%% file: beyond-hypoelasticity-elsarticle.tex
\documentclass[final,3p,authoryear]{elsarticle}

\usepackage{amssymb}
\usepackage{amsmath}

\usepackage{natbib}

\usepackage[english]{babel}
\usepackage[utf8]{inputenc}

\usepackage{subfig}
\usepackage{graphicx}

\usepackage[unicode]{hyperref}

\usepackage[bbgreekl]{mathbbol}
\usepackage{MnSymbol} 

\usepackage{units}
\usepackage{tensor}
\usepackage{accents}

\input{vit-prusa-macros-experimental}

\input{negative-mass-macros}

\usepackage{xcolor}
\usepackage{mdframed}
\usepackage{newfloat}
\input{vit-prusa-summary-style}

\journal{European Journal of Mechanics/A Solids}

\begin{document}

\begin{frontmatter}



\title{Modeling metamaterials by second-order rate-type constitutive relations between only the macroscopic stress and strain\tnoteref{t1}} 

\tnotetext[t1]{Vít Průša thanks the Czech Science Foundation, grant no. 20-11027X, for its support. Casey Rodriguez thanks the NSF, DMS-2307562, for its support. This work has been conceived with the help of late professor K. R. Rajagopal. The authors would like to express gratitude to him for his long-lasting and fruitful collaboration.}


\author[label1]{V\'{\i}t Pr\r{u}\v{s}a\corref{cor1}} 
\ead{prusv@karlin.mff.cuni.cz}

\author[label2]{Casey Rodriguez}
\ead{crodrig@email.unc.edu}

\author[label1]{Ladislav Trnka}
\ead{ladtrnjr@volny.cz}

\author[label1]{Martin Vejvoda}
\ead{martin.vejvoda@mff.cuni.cz}

\cortext[cor1]{Corresponding author.}

\affiliation[label1]{organization={Mathematical Institute, Faculty of Mathematics and Physics, Charles University},
            addressline={Sokolovská~83}, 
            city={Prague},
            postcode={CZ 186\;75}, 
            country={Czech Republic}}
          \affiliation[label2]{
            organization={Department of Mathematics, University of North Carolina},
            addressline={329 Phillips Hall},
            city={Chapel Hill},
            postcode={NC 27599},
            state={North Carolina},
            country={United States of America}
          }

\begin{abstract}
  \input{beyond-hypoelasticity-abstract}
\end{abstract}



\begin{keyword}
  metamaterials \sep mathematical modelling \sep constitutive relations \sep nonlinear elasticity


  \MSC[2008]
  74B20
  \sep
  74B05
  \sep
  74H45
  \sep
  74J05
\end{keyword}

\end{frontmatter}



\input{text/beyond-hypoelasticity-body}

\bibliographystyle{elsarticle-harv} 
\bibliography{vit-prusa,casey-rodriguez}

\end{document}

%% file: vit-prusa-macros-experimental.tex
\DeclareMathOperator{\divergence}{div}

\DeclareMathOperator{\Tr}{Tr}

\newcommand{\exponential}[1]{\ensuremath{{\mathrm e}^{#1}}}

\newcommand{\iunit}{\ensuremath{\mathrm{i}}}

\newcommand{\bydefinition}{\mathrm{def}}

\newcommand{\diff}{\mathrm{d}}

\renewcommand{\vec}[1]{\ensuremath{\mathbf{#1}}}

\makeatletter
\@ifpackageloaded{bm}%
{\renewcommand{\vec}[1]{\ensuremath{\bm{#1}}}%
}{%
\relax
}
\makeatother

\newcommand{\tensorq}[1]{\ensuremath{\mathbb{#1}}}      

\newcommand{\transpose}[1]{#1^\top}

\newcommand{\inverse}[1]{#1^{-1}}

\newcommand{\identity}{\ensuremath{\tensorq{I}}} 
\newcommand{\tensorzero}{{\mathbb{O}}} 

\newcommand{\cstress}{\tensorq{T}}

\newcommand{\thcstress}{\ensuremath{\cstress_{\mathrm{th}}}} 
\newcommand{\thcstressrho}{\ensuremath{\cstress_{\mathrm{th},\, \mathnormal{\rho}}}} 

\newcommand{\fgrad}{\tensorq{F}}

\newcommand{\displacement}{\vec{U}}

\newcommand{\lcg}{\tensorq{B}}

\makeatletter
\@ifpackageloaded{bm}%
{%
\newcommand{\linstrain}{\bbespilon} 
}{%
\newcommand{\linstrain}{\bbespilon} 
}

\@ifpackageloaded{bm}%
{%
 
}{%

}

\@ifpackageloaded{bm}%
{%
\newcommand{\linstress}{\bbtau} 
}{%
\newcommand{\linstress}{\bbtau} 
}
\makeatother

\newcommand{\henckystrain}{\tensorq{H}} 

\newcommand{\generictensor}{{\tensorq{A}}}

\newcommand{\vecv}{\ensuremath{\vec{v}}}
\newcommand{\gradv}{\ensuremath{\nabla \vecv}}
\newcommand{\gradasym}{\ensuremath{\tensorq{W}}}
\newcommand{\gradsym}{\ensuremath{\tensorq{D}}}

\newcommand{\logspin}{\genspin[\mathrm{log}]}
\newcommand{\genspin}[1][\star]{\ensuremath{\tensorq{\Omega}}^{#1}}


\newcommand{\ienergy}{\ensuremath{e}} 
\newcommand{\fenergy}{\ensuremath{\psi}} 
\newcommand{\entropy}{\ensuremath{\eta}} 
\newcommand{\gibbs}{\ensuremath{g}} 

\newcommand{\temp}{\ensuremath{\theta}} 

\newcommand{\rhor}{\rho_{\mathrm{R}}}

\newcommand{\hfluxc}{\vec{j}_{q}}     

\newcommand{\entprodctemp}{\zeta} 

\newcommand{\pd}[2]{\ensuremath{\frac{\partial {#1}}{\partial {#2}}}}
\newcommand{\ppd}[2]{\ensuremath{\frac{\partial^2 {#1}}{\partial {#2^2}}}}
\newcommand{\dd}[2]{\ensuremath{\frac{\diff {#1}}{\diff {#2}}}}

\newcommand{\ddd}[2]{\ensuremath{\frac{\diff^2 {#1}}{\diff {#2}^2}}}

\newcommand{\logfid}[1]{\ensuremath{\accentset{\medcircle_{\mathrm{log}}}{#1}}}
\newcommand{\logfidd}[1]{\ensuremath{\accentset{\medcircle_{\mathrm{log}} \! \medcircle _{\mathrm{log}}}{#1}}}

\makeatletter
\@ifpackageloaded{tensor}
{

}{%

}
\makeatother

\makeatletter
\@ifpackageloaded{tensor}
{

}{%

}
\makeatother

\newcommand{\norm}[2][]{\ensuremath{\left\|#2\right\|_{#1}}}

\makeatletter
\@ifundefined{volume}{%
}%
{%
}
\makeatother

\newcommand{\cvolumee}{\diff \mathrm{v}}

\newcommand{\tensortensor}[2]{\ensuremath{#1 \otimes #2}}

\makeatletter

\@ifpackageloaded{MnSymbol} 
{
\newcommand{\tensordot}[2]{\ensuremath{#1 \vdotdot #2}} 
}{%
\newcommand{\tensordot}[2]{\ensuremath{#1 : #2}} 
}

\@ifpackageloaded{MnSymbol} 
{
   
}{%
   
}
\makeatother

\newcommand{\vectordot}[2]{\ensuremath{#1 \bullet #2}}



%% file: negative-mass-macros.tex
\newcommand{\meff}{m_{\text{eff}}}

\newcommand{\difference}{\text{diff}}

\newcommand{\cstressrho}{\ensuremath{\cstress_{\mathnormal{\rho}}}} 
\newcommand{\thcstressDiff}{\ensuremath{\cstress_{\mathrm{th},\, \mathrm{diff}}}} 
\newcommand{\thcstressrhoDiff}{\ensuremath{\cstress_{\mathrm{th},\, \mathnormal{\rho}_2,\, \mathrm{diff}}}} 

\newcommand{\thcstressrhoA}{\ensuremath{\cstress_{\mathrm{th},\, \mathnormal{\rho}_1,\, \mathrm{1}}}} %

\newcommand{\thcstressA}{\ensuremath{\cstress_{\mathrm{th},\, \mathrm{1}}}} %

\newcommand{\henckystrainA}{\henckystrain_{\mathrm{1}}} 
\newcommand{\henckystrainB}{\henckystrain_{\mathrm{2}}} 
\newcommand{\henckystrainDiff}{\henckystrain_{\mathrm{diff}}} 

\newcommand{\linstrainA}{\linstrain_{\mathrm{1}}} 
\newcommand{\linstrainB}{\linstrain_{\mathrm{2}}} 
\newcommand{\linstrainDiff}{\linstrain_{\mathrm{diff}}} 

\newcommand{\linstressA}{\linstress_{\mathrm{1}}} 
\newcommand{\linstressDiff}{\linstress_{\mathrm{diff}}} 

\newcommand{\gibbsB}{\mathit{\gibbs}_{\mathrm{2}}} 
\newcommand{\fenergyA}{\mathit{\fenergy}_{\mathrm{1}}} 
\newcommand{\fenergyB}{\mathit{\fenergy}_{\mathrm{2}}} 

\newcommand{\discstressrho}{\ensuremath{\cstress_{\mathrm{diss}, \mathnormal{\rho}}}} 

\newcommand{\foidentity}{\ensuremath{\fotensorq{I}}} 
\newcommand{\fotensorq}[1]{\mathcal{#1}} 

\newcommand{\stiffnesstensor}{\ensuremath{\fotensorq{C}}} 
\newcommand{\stiffnesstensorDiff}{\ensuremath{\fotensorq{C}_{\mathrm{diff}}}} 
\newcommand{\ratestiffnesstensor}{\ensuremath{\fotensorq{C}_{\mathrm{rate}}}} 

\newcommand{\rhoA}{\rho_{\mathrm{1}}} 
\newcommand{\rhoB}{\rho_{\mathrm{2}}} 
\newcommand{\rhorB}{\rho_{\mathrm{2}, \mathrm{R}}} 

%% file: vit-prusa-summary-style.tex
\DeclareFloatingEnvironment[fileext=sum,placement={!ht},name=Summary]{summaryflt}

\newenvironment{summary}[1][]
    {
    \begin{summaryflt}[tb]
        \begin{summarycont}[#1] 
    }
    {
        \end{summarycont}
        \end{summaryflt}
    }

\mdfdefinestyle{summarystyle}{%
linecolor=black,linewidth=3pt,%
frametitlerule=true,%
frametitlebackgroundcolor=gray!20,
innertopmargin=\topskip,
}

\mdtheorem[style=summarystyle]{summarycont}{Summary}

\newmdenv[style=scholionstyle]{scholion}
\mdfdefinestyle{scholionstyle}{
  skipabove=1em,
  skipbelow=1em,
  backgroundcolor=gray!10,
  roundcorner=10pt
}

%% file: beyond-hypoelasticity-abstract.tex
We propose a thermodynamically based approach for constructing effective rate-type constitutive relations describing finite deformations of metamaterials. The effective constitutive relations are formulated as \emph{second-order} in time rate-type Eulerian constitutive relations between only the Cauchy stress tensor, the Hencky strain tensor and objective time derivatives thereof. In particular, there is no need to introduce additional quantities or concepts such as ``micro-level deformation'',``micromorphic continua'', ``enriched continua'', or elastic solids with frequency dependent material properties. The linearisation of the proposed fully nonlinear (finite deformations) constitutive relations leads, in Fourier space, to the same constitutive relations as those commonly used in theories based on the concepts of frequency dependent density and/or stiffness. From this perspective the proposed constitutive relations reproduce the behaviour predicted by the frequency dependent density and/or stiffness models, but yet they work with constant---that is motion independent---material properties. Finally, we argue that the proposed fully nonlinear (finite deformations) second-order in time rate-type constitutive relations do not fall into traditional classes of models for elastic solids (hyperelastic solids/Green elastic solids, first-order in time hypoelastic solids), and that the proposed constitutive relations embody a \emph{new} class of constitutive relations characterising elastic solids.


%% file: text/beyond-hypoelasticity-body.tex
\section{Introduction}
\label{sec:introduction}

\begin{figure}[h]
  \centering
  \subfloat[\label{fig:internal-arrangement-a} Material filled with densely arranged microcavities, {\cite[Figure 2]{milton.gw.willis.jr:on}}.]{\includegraphics[width=0.35\textwidth]{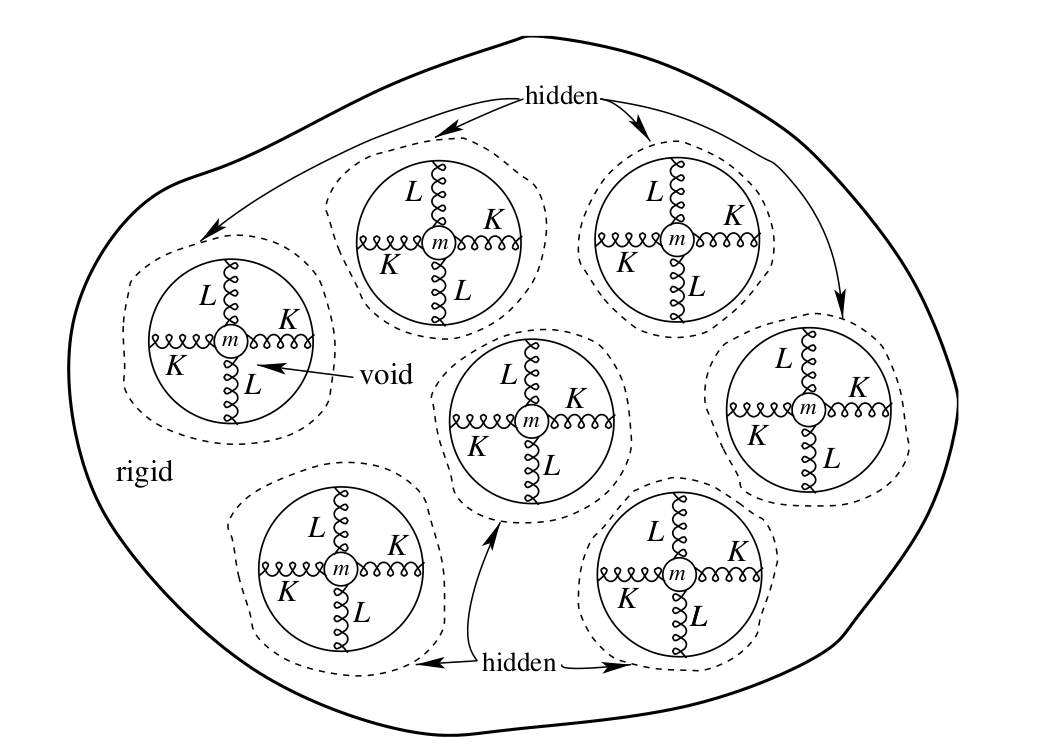}}
  \quad
  \subfloat[\label{fig:internal-arrangement-b} Elementary cell in a three component composite material, {\cite[Figure 1]{liu.z.chan.ct.ea:analytic}}.]{\includegraphics[width=0.26\textwidth]{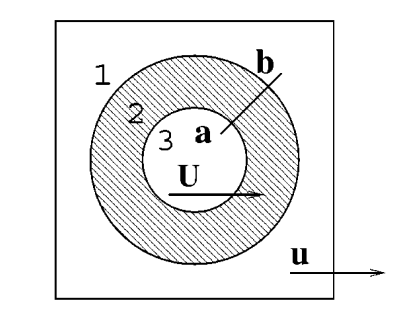}}
  \quad
  \subfloat[\label{fig:internal-arrangement-c} Elementary cell in a microstructured material, {\cite[Figure 2b]{lee.j.kim.yy:elastic}}.]{\includegraphics[width=0.35\textwidth]{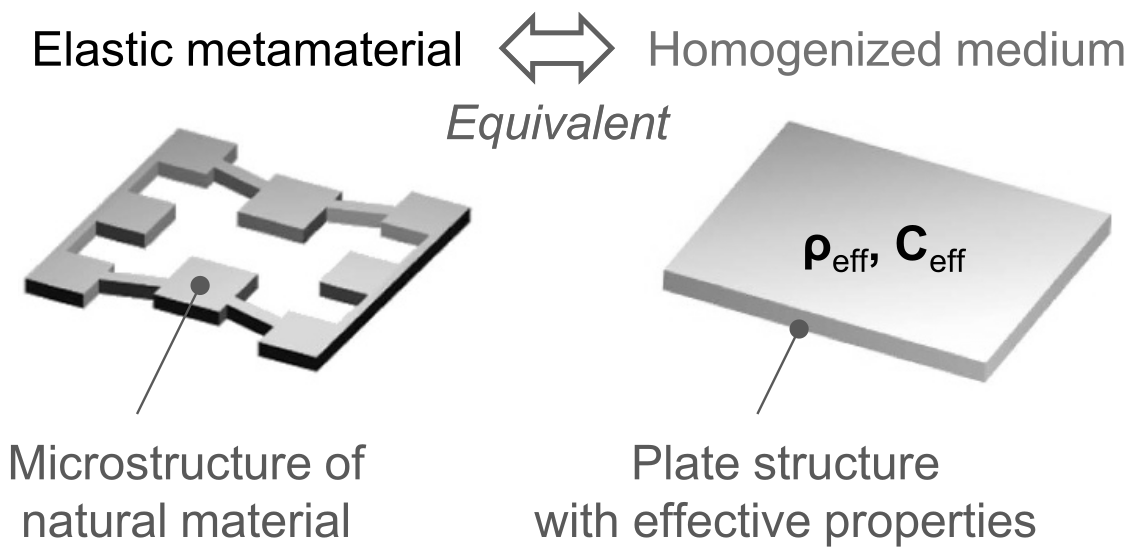}}
  \caption{Schematics of internal arrangement of metamaterials.}
  \label{fig:internal-arrangement}
\end{figure}

The term \textit{metamaterial} is typically used for an artificial composite material with a carefully engineered microstructure. Most interesting microstructures have a surprising impact on the overall macroscopic dynamical behaviour of such a material despite the heterogeneous microstructure itself not being directly observable at the macroscopic level. For example, materials formed by periodically arranged elementary cells, see Figure~\ref{fig:internal-arrangement} for schematics of some arrangements, might have surprising properties from the perspective of \emph{wave propagation}, see~\cite{lee.j.kim.yy:elastic} or \cite{rizzi.g.d’agostino.mv.ea:from} for numerous examples and further references. Due to its unique properties, the study of metamaterials is a subject of intensive research activity, see~\cite{failla.g.marzani.a.ea:current}. 

The macroscopic behaviour of metamaterials is frequently described using the \emph{effective mass concept}, see, for example, \cite{chan.ct.li.j.ea:on}, \cite{li.j.chan.ct:double-negative}, \cite{liu.z.chan.ct.ea:analytic}, \cite{sheng.p.zhang.xx.ea:locally}, \cite{milton.gw.willis.jr:on} and \cite{huang.hh.sun.ct.ea:on}, to name a few early works based on this idea. Using the effective mass concept the material of interest is---from the perspective of wave propagation---described as a standard (linearised) elastic solid with \emph{frequency dependent} mass/density and/or stiffness.

\begin{figure}[h]
  \centering
  \subfloat[\label{fig:mass-in-mass-a}Fully resolved system with visible inner mass $m_2$.]{\includegraphics[width=0.3\textwidth]{figures/mass-in-mass}}
  \qquad
  \subfloat[\label{fig:mass-in-mass-b}Effective single mass system, effective mass~$\meff$.]{\includegraphics[width=0.3\textwidth]{figures/mass-in-mass-effective}}
  \qquad
  \subfloat[\label{fig:mass-in-mass-c}Effective single mass system, effective linear constitutive relation $\sigma = \mathcal{L}(x_1)$.]{\includegraphics[width=0.3\textwidth]{figures/mass-in-mass-effective-spring}}
  \caption{Mass-in-mass model.}
  \label{fig:mass-in-mass}
\end{figure}

The concepts of \emph{frequency} dependent effective mass/density and stiffness are based on detailed analyses of simple mechanical systems, see~\cite{milton.gw.willis.jr:on} and \cite{huang.hh.sun.ct.ea:on}. In particular, \cite{huang.hh.sun.ct.ea:on} motivate the concept of frequency dependent density/mass via the analysis of the simple mass-in-mass system shown in Figure~\ref{fig:mass-in-mass}. Having a fully resolved system with an ``internal resonator'', see Figure~\ref{fig:mass-in-mass-a}, the objective is to develop an effective model, see Figure~\ref{fig:mass-in-mass-b}, that correctly predicts the dynamic behaviour of the fully resolved mass-in-mass system, but without a direct reference to the motion of the ``internal resonator'' represented by the inner mass $m_2$ and the inner spring with stiffness $k_2$. If the system is subject to periodic forcing, \cite{milton.gw.willis.jr:on} and~\cite{huang.hh.sun.ct.ea:on} achieve this goal by introducing the \emph{frequency dependent effective mass} $\meff$. However, this frequency dependent effective mass is a quantity that makes sense only in Fourier space. Similar ideas are then transferred to lattice systems composed of mass-in-mass elementary cells and finally to the linearised elasticity setting. In the linearised elasticity setting, the concept of frequency dependent effective mass density and/or stiffness manifests itself in the \emph{Fourier transformed} version of the governing equations that read
\begin{subequations}
  \label{eq:122}
  \begin{align}
    \label{eq:134}
    - \widehat{\rho}(\omega) \omega^2 \widehat{\displacement} &= \widehat{\divergence \linstress}, \\
    \label{eq:135}
    \widehat{\linstress}
                                &=
                                  \widehat{\stiffnesstensor}(\omega)
     \widehat{\linstrain},
  \end{align}
\end{subequations}
where $\omega$ denotes the wave frequency, $\displacement$ denotes the displacement, $\linstress$ denotes the stress, $\linstrain$ denotes the linearised strain tensor, $\widehat{\rho}(\omega)$ denotes the frequency dependent density and $\widehat{\stiffnesstensor}(\omega)$ denotes the frequency dependent stiffness tensor. Here, a hat denotes the object's spacetime Fourier transform. If $\widehat{\rho}$ and $\widehat{\stiffnesstensor}$ are constant (frequency independent), then the system~\eqref{eq:122} is identical to the (Fourier image of) governing equations in the standard linearised elasticity case,
\begin{subequations}
  \label{eq:159}
  \begin{align}
    \label{eq:160}
    \rho
    \ppd{\displacement}{t}
    &=
      \divergence \linstress
      ,
    \\
    \label{eq:165}
      \linstress 
    &=
      \stiffnesstensor \linstrain
      ,
  \end{align}
\end{subequations}
wherein $\stiffnesstensor$ and $\rho$ are constants.

While the frequency dependent density and/or stiffness concept might be useful in characterizing wave propagation in metamaterials, it has obvious physical inconsistencies. It turns out that the effective density might be, in certain frequency ranges, negative, and in the physical space the frequency dependent density reduces to a \emph{nonlocal density operator}, see~\cite{milton.gw.willis.jr:on}. Similar observations hold for a frequency dependent stiffness. Such an interpretation is at odds with a fundamental principle of classical Newtonian physics---the mass is a constant that does not depend on the state of motion. This contradiction suggests that the concept of effective density/stiffness is simply a convenient mathematical trick.

The unease with frequency dependent density is articulated by \cite{rizzi.g.d’agostino.mv.ea:from}, \cite{smejkal.m.jirasek.m.ea:integral} and \cite{cichra.d.prusa.v.ea:conclusion}, to name a few recent works. The question is how to mitigate the issue. Since the response of metamaterials to wave motion depends on the  wave \emph{frequency}, one can hardly capture this phenomenon via traditional hyperelastic/Green elastic models that are rate-independent. The classical hyperelasticity theory must be transgressed. \cite{rizzi.g.d’agostino.mv.ea:from} and \cite{smejkal.m.jirasek.m.ea:integral} address the problem of frequency dependent material properties by introducing the concept of \emph{micromorphic} material. While the details differ in the aforementioned works, they share the same key idea. The frequency dependent material properties are eliminated by using the concept of ``micro-level deformation'' or ``micromorphic continua'' or ``enriched continua''. In essence, these theories supplement the standard kinematic variables by an additional kinematic variable that effectively captures the behaviour of the ``internal resonator''. Such generalised continua have origins in the works \cite{Piola1846Book} and \cite{Cosserats1909} with modern theories being developed beginning in the 1960's; see, for example, \cite{Toupin62}, \cite{Mindlin64a}, \cite{Toupin64}, \cite{GreenRivlin64b}, \cite{GreenRivlin64a}, \cite{Mindlin1965}, \cite{MindlinEshel1968}, \cite{Germain73a}, \cite{Germain73b} and the reviews \cite{askes.h.aifantis.ec:gradient}, \cite{dellIsola17}, \cite{Maugin17Book}, \cite{dellIsola2020higher}.

The work by~\cite{cichra.d.prusa.v.ea:conclusion} takes an altogether different approach. The authors of \cite{cichra.d.prusa.v.ea:conclusion} revisit the simple mass-in-mass system by~\cite{huang.hh.sun.ct.ea:on}, and they show that the concept of effective mass can be eliminated using the concept of effective constitutive relation. A schematic representation thereof means going from Figure~\eqref{fig:mass-in-mass-a} to Figure~\eqref{fig:mass-in-mass-c}. The effective constitutive relation proposed by~\cite{cichra.d.prusa.v.ea:conclusion} is a \textit{second-order} rate-type constitutive relation between the displacement/strain and the force/stress only; it \emph{neither} directly refers to the ``internal resonator'' \emph{nor} does it introduce some additional concept (such as micromorphic continua) beyond the macroscopic force/stress and the displacement/strain.

The analysis in~\cite{cichra.d.prusa.v.ea:conclusion} however deals only with the simple mass-in-mass system and a lattice model composed of these simple mass-in-mass systems. In the present work, we revisit the work by~\cite{cichra.d.prusa.v.ea:conclusion}, and we develop an approach that allows one to design a class of fully nonlinear continuum models for finite deformations that have---from the perspective of small amplitude wave propagation---the same properties as the popular frequency dependent effective density/stiffness models~\eqref{eq:122}. The fully nonlinear second-order rate-type effective constitutive relation we propose reads
\begin{subequations}
  \label{eq:137}
  \begin{align}
    \label{eq:139}
    -
    \ratestiffnesstensor
    \logfidd{
      \overline{
        \left(
          \henckystrain
          +
    \pd{\gibbsB}{\thcstressrhoDiff}
        \right)
      }
    }
    +
    \thcstressrhoDiff
    &=
    \tensorzero
      ,
      \\
    \label{eq:151}
    \thcstressrhoDiff &= \cstressrho - \pd{\fenergyA}{\henckystrain}.
  \end{align}
\end{subequations}
(The effective constitutive relation is formulated in the Eulerian description.) Here, the symbol $\cstressrho =_{\bydefinition} \frac{\cstress}{\rho}$ denotes the Cauchy stress tensor divided by the current apparent material density~$\rho$, $\henckystrain$ denotes the Hencky strain tensor and $\ratestiffnesstensor$ is a positive definite fourth order tensor. The objective derivative in~\eqref{eq:139} is the logarithmic corotational derivative, see~\cite{xiao.h.bruhns.ot.ea:logarithmic}. A particular constitutive relation is obtained by the choice of two potentials, $\gibbsB$ and $\fenergyA$. These potentials mimic the Helmholtz free energy/Gibbs free energy characterising the mechanical response of the matrix material and the ``internal resonators''.  We emphasize that~\eqref{eq:137} works only with the (overall) Cauchy stress tensor~$\cstress$ and the Hencky strain tensor~$\henckystrain$, no microscopic kinematic variable is introduced.

\emph{If the nonlinear constitutive relations~\eqref{eq:137} are linearised}, then the full system of governing equations \emph{in the physical domain}, including the balance of linear momentum, reads
\begin{subequations}
  \label{eq:152}
  \begin{align}
    \label{eq:153}
    \rhor
    \ppd{\displacement}{t}
    &=
      \divergence \linstress
      ,
    \\
    \label{eq:154}
    -
    \ratestiffnesstensor
    \rhor
    \ppd{}{t}
    \left(
    \linstrain
    -
    \frac{\rhorB}{\rhor}
    \inverse{\stiffnesstensorDiff} 
    \left(
      \linstress - \stiffnesstensor \linstrain
    \right)
  \right)
    +
    \left(
      \linstress - \stiffnesstensor \linstrain
    \right)
    &=
    \tensorzero,
  \end{align}
\end{subequations}
where $\rhor$ and $\rhorB$ denote the (constant) material densities in the reference configuration, $\ratestiffnesstensor$, $\stiffnesstensorDiff$ and~$\stiffnesstensor$ denote (constant) stiffness-like fourth order tensors, and $\linstrain$ denotes the standard linearised strain tensor. The Fourier transformed version of~\eqref{eq:152} has, upon the appropriate choice of material constants, the same properties as~\eqref{eq:122}. In particular, the Fourier transformed version of~\eqref{eq:154} can be formally rewritten in the form~\eqref{eq:135}, and hence, the model has---from the perspective of wave propagation---the same properties of the frequency dependent effective density/stiffness model~\eqref{eq:122}.

The key feature of the newly proposed model~\eqref{eq:152} compared to frequency-dependent materials parameters model~\eqref{eq:122} is that~\eqref{eq:152} is formulated and applies in the \emph{physical} domain. This means that the model~\eqref{eq:152} based on the concept of~\emph{effective constitutive relation} can be, unlike the model~\eqref{eq:152}, utilized to study phenomena beyond wave propagation in an infinite domain. \emph{Furthermore, the effective system~\eqref{eq:152} is just a linearisation of a fully nonlinear model~\eqref{eq:137} for finite deformations, and the existence of a fully nonlinear model behind~\eqref{eq:154} suggests that the proposed approach may also provide the tools to investigate genuine nonlinear phenomena in metamaterials.} 

A brief summary of our work is as follows. In Section~\ref{sec:simple-one-dimens}, we first illustrate the basic ideas behind the fully nonlinear continuous rate-type model~\eqref{eq:137} via a careful study of the paradigmatic simple one-dimensional discrete mass-in-mass model for a metamaterial behaviour, see~\cite{huang.hh.sun.ct.ea:on}. In Section~\ref{sec:linear}, we first consider a mass-in-mass system with \emph{linear} springs, and we recall the basic concept of \emph{effective constitutive relation} used in analysis by~\cite{cichra.d.prusa.v.ea:conclusion}. Next, we proceed with the one-dimensional discrete mass-in-mass model with \emph{nonlinear} springs in Section~\ref{sec:non-linear}. In particular, we study the interplay between the \emph{nonlinear} spring forces, the corresponding potentials and the final effective rate-type constitutive relation. The analysis of the simple discrete one-dimensional models provides us a motivation for the construction of the fully nonlinear rate-type model \eqref{eq:137} for finite deformation of a continuous medium (metamaterial). The construction itself is described in Section~\ref{sec:fully-three-dimens}. In Section \ref{sec:dispersion-relation}, we then derive the linearized system \eqref{eq:152}, and we show that the associated dispersion relation has the same form as the dispersion relation obtained in metamaterial models that use the concept of frequency dependent density. Our conclusions and comparison of the proposed model to certain traditional models of elastic solids is presented in Section~\ref{sec:conclusion}.

\section{Simple one-dimensional discrete mass-in-mass system}
\label{sec:simple-one-dimens}

In this section, we revisit the simple one-dimensional discrete mass-in-mass model by~\cite{huang.hh.sun.ct.ea:on}, and we show how to obtain an effective rate-type constitutive relation describing the mechanical response of such a system. This analysis allows us to identify the foundational concepts that lead to the fully nonlinear continuous rate-type model~\eqref{eq:137}.

\subsection{Mass-in-mass system with linear springs}
\label{sec:linear}
We first consider the discrete mass-in-mass system with \textit{linear} springs, see Figure~\eqref{fig:mass-in-mass-a}, governed by the equations
\begin{subequations}
  \label{eq:9}
  \begin{align}
    \label{eq:10}
    m_1 \ddd{x_1}{t}  + k_1 x_1 + k_2 \left(x_1 - x_2\right) & = f, \\
    \label{eq:11}
    m_2 \ddd{x_2}{t}  - k_2 \left(x_1 - x_2\right) & = 0,
  \end{align}
\end{subequations}
see, for example, \cite{huang.hh.sun.ct.ea:on}. \cite{cichra.d.prusa.v.ea:conclusion} have shown that the governing equations for the simple discrete mass-in-mass system~\eqref{eq:9} can be rewritten in the form
\begin{subequations}
  \label{eq:107}
  \begin{align}
    \label{eq:108}
    m_1 \ddd{x_1}{t}  - \sigma &= f, \\
    \label{eq:109}
    \ddd{\sigma}{t} + \frac{k_2}{m_2} \sigma &= -\left(k_1 + k_2\right) \ddd{x_1}{t} - \frac{k_1 k_2}{m_2} x_1,
  \end{align}
\end{subequations}
where
\begin{equation}
  \label{eq:110}
  \sigma = _{\bydefinition}  -  k_1 x_1 - k_2 \left(x_1 - x_2\right).
\end{equation}
In the rest of the present work we refer to the system~\eqref{eq:107} as the \emph{effective system of governing equations} for the simple discrete mass-in-mass system (with linear springs), and the equation~\eqref{eq:109} is called the \emph{effective constitutive relation}. We note that the effective constitutive relation~\eqref{eq:109} can be also rewritten as 
\begin{equation}
  \label{eq:112}
  \dd{}{t}
  \left(
    \dd{}{t}
    \left(
      x_1 + \frac{\sigma + k_1x_1}{k_2}
    \right)
  \right)
  +
  \frac{\sigma + k_1x_1}{m_2}
  =
  0
  ,
\end{equation} 
which is the formula we later generalise into the fully nonlinear continuous three-dimensional setting.

The systems of governing equations~\eqref{eq:9} and~\eqref{eq:107} are equivalent, and if the external forcing is equal to zero, $f=0$, then both systems of equations possess a conserved quantity---the energy. In this sense both systems serve as simple models for a hyperelastic/Green elastic solid. The total energy~$E$---that is the kinetic energy plus the elastic stored energy---for the discrete mass-in-mass system~\eqref{eq:9} reads
\begin{equation}
  \label{eq:111}
  E
  =_{\bydefinition}
  \frac{1}{2}
  m_1
  \left(
    \dd{x_1}{t}
  \right)^2
  +
  \frac{1}{2}
  m_2
  \left(
    \dd{x_2}{t}
  \right)^2
  +
  k_1
  \frac{x_1^2}{2}
  +
  k_2 \frac{\left(x_1 - x_2\right)^2}{2}
  .
\end{equation}
The meaning of the individual terms in the energy formula is clear---they are in order the kinetic energy of the outer mass, the kinetic energy of the inner mass, the elastic stored energy of the outer linear spring and the elastic stored energy of the inner linear spring.

As observed in~\cite{cichra.d.prusa.v.ea:conclusion}, we can rewrite the energy~\eqref{eq:111} in terms of natural variables for the effective system~\eqref{eq:107}, and doing so we get
\begin{equation}
  \label{eq:113}
  E
  =
  \frac{1}{2}
  m_1
  \left(
    \dd{x_1}{t}
  \right)^2
  +
  \underbrace{
  \frac{1}{2}
  \frac{m_2}{k_2^2}
  \left(
    \dd{\sigma}{t} + \left(k_1 + k_2\right) \dd{x_1}{t}
  \right)^2
  +
  k_1
  \frac{x_1^2}{2}
  +
  \frac{\left( \sigma +  k_1 x_1\right)^2}{2k_2}
}_W
.
\end{equation}
The first term is the kinetic energy of the outer mass, and \emph{the collection of the remaining terms} denoted by $W$ \emph{is interpreted as a generalised elastic stored energy} for a virtual elastic material described by the effective constitutive relation~\eqref{eq:112}. Naturally, one of the terms in the newly identified generalised elastic stored energy $W$ is the elastic stored energy of the outer mass~$k_1\frac{x_1^2}{2}$, but there are other contributions. Unlike in the classical case wherein the elastic stored energy is a function of displacement/strain only, these new contributions depend on the stress/force~$\sigma$, displacement/strain~$x_1$ and the time derivatives of these quantities. \emph{The interpretation of the corresponding terms as a generalised elastic stored energy constitutes the key step in the later construction of the fully nonlinear (finite deformations) rate-type model for continuous medium representing the metamaterial.}



\subsubsection{From the energy to the effective constitutive relation}
\label{sec:from-energy-effect-1}
We now show how to obtain the effective model~\eqref{eq:107} with the constitutive relation~\eqref{eq:109} directly from the generalised elastic stored energy $W$ given in~\eqref{eq:113}. As discussed above, the conserved total energy for the system is
\begin{equation}
  \label{eq:1}
    E
  =
  \frac{1}{2}
  m_1
  \left(
    \dd{x_1}{t}
  \right)^2
  +
  \frac{1}{2}
  \frac{m_2}{k_2^2}
  \left(
    \dd{\sigma}{t} + \left(k_1 + k_2\right) \dd{x_1}{t}
  \right)^2
  +
  k_1
  \frac{x_1^2}{2}
  +
  \frac{\left( \sigma +  k_1 x_1\right)^2}{2k_2},
\end{equation}
see~\eqref{eq:113}, and we identify the last three terms as the generalised elastic stored energy
\begin{equation}
  \label{eq:5}
  W = W \left( x_1, \dd{x_1}{t}, \sigma, \dd{\sigma}{t} \right).
\end{equation}
The generalised elastic stored energy now depends on the displacement/strain and the force/stress and \emph{their time derivatives}---this contrasts to the classical concept of elastic stored energy which is assumed to be a function of the displacement/strain only. We thus set
\begin{equation}
  \label{eq:2}
  W
  =_{\bydefinition}
  \frac{1}{2}
  \frac{m_2}{k_2^2}
  \left(
    \dd{\sigma}{t} + \left(k_1 + k_2\right) \dd{x_1}{t}
  \right)^2
  +
  k_1
  \frac{x_1^2}{2}
  +
  \frac{\left( \sigma +  k_1 x_1\right)^2}{2k_2}.
\end{equation}
Now we introduce a procedure that allows us to identify the effective constitutive relation implied by this choice of this generalised elastic stored energy. In principle, our method follows the classical procedure that in the standard continuum mechanics elasticity theory leads to the relation between a potential (elastic stored energy, strain-energy) and the stress; see, for example, \cite{truesdell.c.noll.w:non-linear*1}.

In continuum mechanics the evolution equation for the internal energy in the absence of heat flux reads
$
  \rho \dd{\ienergy}{t}
  =
  \tensordot{
    \cstress
  }
  {
    \gradsym
  }
  $. (This equation is also called the stress-power formula.) Assuming that $m_1 \ddd{x}{t} - \sigma = f$, the discrete analogue of the stress-power formula
  $
  \rho \dd{\ienergy}{t}
  =
  \tensordot{
    \cstress
  }
  {
    \gradsym
  }
$ is the equation
\begin{equation}
  \label{eq:4}
  \dd{W}{t} = - \sigma \dd{x_1}{t}.
\end{equation}
 We assume that the generalised elastic stored energy is given by~\eqref{eq:2} and proceed. We use the chain rule on the left-hand side of~\eqref{eq:4}, and~\eqref{eq:4} then in fact produces the evolution equation for the entropy in the discrete setting; see Section~\ref{sec:fully-three-dimens} for details in the three-dimensional setting. We identify the stress-strain relation/constitutive relation, that is the relation between~$\sigma$ and $x_1$, via the requirement that the ``entropy production'' vanishes in all mechanical processes, that is we choose the stress--strain relation such that~\eqref{eq:4} holds.

The chain rule, with the usual abuse of notation, yields
\begin{multline}
  \label{eq:6}
  \dd{W}{t}
  =
  \pd{W}{x_1}
  \dd{x_1}{t}
  +
  \pd{W}{\dd{x_1}{t}}
  \ddd{x_1}{t}
  +
  \pd{W}{\sigma}
  \dd{\sigma}{t}
  +
  \pd{W}{\dd{\sigma}{t}}
  \ddd{\sigma}{t}
  \\
  =
  k_1
  x_1
  \dd{x_1}{t}
  +
  \frac{\sigma +  k_1 x_1}{k_2}
  \left(
    \dd{\sigma}{t}
    +
    k_1
    \dd{x_1}{t}
  \right)
  +
  \frac{m_2}{k_2^2}
  \left(
    \dd{\sigma}{t} + \left(k_1 + k_2\right) \dd{x_1}{t}
  \right)
  \left(
    \ddd{\sigma}{t} + \left(k_1 + k_2\right) \ddd{x_1}{t}
  \right)
  .
\end{multline}
Now we use~\eqref{eq:6} on the left-hand side of~\eqref{eq:4}, and upon rearranging the terms we arrive at
\begin{equation}
  \label{eq:7}
  0
  =
  \left[
    \dd{\sigma}{t} + \left(k_1 + k_2\right) \dd{x_1}{t}
  \right]
  \left[
    \frac{\sigma +  k_1 x_1}{k_2}
    +
    \frac{m_2}{k_2^2}
    \left(
      \ddd{\sigma}{t} + \left(k_1 + k_2\right) \ddd{x_1}{t}
    \right)
  \right]
  .
\end{equation}

We assume that at any particular time, the values of $\dd{\sigma}{t}$ and $\dd{x_1}{t}$ can be arbitrarily assigned independently of each other. This assumption aligns with the form of $\sigma$ in \eqref{eq:110}, as describing the motion of the internal mass necessitates prescribed initial conditions for its position and velocity, which are independent of those of the outer mass. Analogous assumptions about the deformation gradient and the entropy appear in the classical setting of three-dimensional continuous theories of elasticity; see \cite{coleman.bd.noll.w:thermodynamics} or \cite{truesdell.c.noll.w:non-linear*1}. Thus, if we want~\eqref{eq:7} to be identically satisfied in all mechanical processes, which means that the entropy production is equal to zero in all mechanical processes, then we must fix the second bracket to be equal to zero, which yields
\begin{equation}
  \label{eq:8}
  \ddd{\sigma}{t} + \frac{k_2}{m_2} \sigma = -\left(k_1 + k_2\right) \ddd{x_1}{t} - \frac{k_1 k_2}{m_2} x_1.
\end{equation}
Equation~\eqref{eq:8} is the effective constitutive relation~\eqref{eq:109}, identified in~\cite{cichra.d.prusa.v.ea:conclusion}; see also~\eqref{eq:112} for an alternative form of the effective constitutive relation.

We observe that~\eqref{eq:7} hints at how to prove the conservative property of system~\eqref{eq:107} with $f=0$. In particular, we take the effective constitutive relation~\eqref{eq:109}, and we multiply it by $\dd{\sigma}{t} + \left(k_1 + k_2\right) \dd{x_1}{t}$, which gives us, after some algebra, the equation~\eqref{eq:4}. Furthermore, we multiply~\eqref{eq:108} with $\dd{x_1}{t}$, which yields
\begin{equation}
  \label{eq:106}
  \dd{}{t} \left(  \frac{1}{2} m_1 \left( \dd{x_1}{t} \right)^2 - \sigma \dd{x_1}{t} \right) = 0,
\end{equation}
and substituting~\eqref{eq:4} into~\eqref{eq:106} then yields
\begin{equation}
  \label{eq:114}
  \dd{}{t} \left(  \frac{1}{2} m_1 \left( \dd{x_1}{t} \right)^2 + W \right) = 0,
\end{equation}
which shows that $\frac{1}{2} m_1 \left( \dd{x_1}{t} \right)^2 + W$ is a conserved quantity as requested.

\subsection{Mass-in-mass system with non-linear springs}
\label{sec:non-linear}
Having analysed the \emph{linear} discrete mass-in-mass system, we can proceed with the~\emph{nonlinear} discrete mass-in-mass system. Since \cite{cichra.d.prusa.v.ea:conclusion} have dealt with \emph{linear} springs only, we first need to derive the effective constitutive relation for the fully resolved one-dimensional mass-in-mass model with \emph{nonlinear} springs. 

\subsubsection{From the fully resolved model to the effective model }
\label{sec:from-fully-resolved}
We start with a nonlinear version of the model mass-in-mass system~\eqref{eq:9}. If the springs are nonlinear, then a variant of~\eqref{eq:9} reads
\begin{subequations}
  \label{eq:13}
  \begin{align}
    \label{eq:14}
    m_1 \ddd{x_1}{t} - \sigma_1 + \sigma_{\difference} & = f, \\
    \label{eq:15}
    m_2 \ddd{x_2}{t} - \sigma_{\difference} & = 0,
  \end{align}
\end{subequations}
where $\sigma_{\difference}$ denotes the force/stress exerted by the inner mass to the outer mass, and $\sigma_1$ denotes the stress exerted by the outer/first spring to the outer mass. 

Now we make the key assumption. We assume that the outer/first spring is a hyperelastic/Green elastic spring with the Helmholtz free energy $\fenergy_1$, while the inner/second spring is a hyperelastic/Green elastic spring with the Gibbs free energy~$\gibbs_2$,
\begin{subequations}
  \label{eq:16}
  \begin{align}
    \label{eq:17}
    \sigma_1 &= - \pd{\fenergy_1}{x_1}, \\
    \label{eq:18}
    x_{\difference} &=_{\bydefinition} x_2 - x_1, \\
    \label{eq:19}
    x_{\difference} &= - \pd{\gibbs_2}{\sigma_{\difference}}.
  \end{align}
\end{subequations}
(Here, we use the sign convention favourable for spring systems, but later, see Section~\ref{sec:fully-three-dimens}, we switch to the sign convention used in continuum mechanics.) The Gibbs and Helmholtz free energies are linked by the Legendre transformation, that is if we start with the Helmholtz free energy with
\begin{equation}
  \label{eq:20}
  - \sigma_1 = \pd{\fenergy_1}{x_1},
\end{equation}
then the corresponding dual potential/Gibbs free energy reads
\begin{equation}
  \label{eq:29}
  \gibbs_1(-\sigma_1) = \left. \left[ \left( - \sigma_1 \right) x_1 - \fenergy_1(x_1) \right] \right|_{x_1 = \inverse{\left( \pd{\fenergy_1}{x_1} \right)}(-\sigma_1)}
\end{equation}
and similarly for the inner/second spring.

The use of the Gibbs free energy for the inner/second spring helps us to avoid the need to explicitly write down formulae for some inverse functions; in other words, the algebra is a little bit simpler if the inner/second spring is characterised by the Gibbs free energy instead of the standard Helmholtz free energy. This is not surprising since all we need to know concerning the inner/second spring is its~\emph{force/stress contribution} to the evolution equation~\eqref{eq:14} for the outer/first mass. Thus, we use the force/stress contribution as the primitive variable rather than the corresponding displacement/strain. This observation is trivial for the one-dimensional system, but it makes the later derivation of fully three-dimensional model much more transparent.

We are now ready to proceed with the derivation of the effective constitutive relations. Let $\sigma$ denote the total force/stress acting on the outer/first mass,
\begin{equation}
  \label{eq:146}
  \sigma =_{\bydefinition} \sigma_1 - \sigma_{\difference}.
\end{equation}
Using this notation the evolution equation for the outer/first mass~\eqref{eq:14} reads
\begin{equation}
  \label{eq:25}
  m_1 \ddd{x_1}{t} - \sigma  = f.
\end{equation}
Differentiation of~\eqref{eq:146} with respect to time yields
\begin{subequations}
  \label{eq:26}
  \begin{align}
    \label{eq:27}
    \dd{\sigma}{t} &= - \ppd{\fenergy_1}{x_1} \dd{x_1}{t} - \dd{\sigma_{\difference}}{t}, \\
    \label{eq:28}
    \ddd{\sigma}{t} &= - \pd{^3\fenergy_1}{x_1^3} \left(\dd{x_1}{t}\right)^2 - \ppd{\fenergy_1}{x_1} \ddd{x_1}{t} - \ddd{\sigma_{\difference}}{t},
  \end{align}
\end{subequations}
where we have used the constitutive relation~\eqref{eq:17} for the outer/first spring. Differentiating the constitutive relation for the inner/second spring~\eqref{eq:19} gives
\begin{subequations}
  \label{eq:30}
  \begin{align}
    \label{eq:31}
    \dd{x_{\difference}}{t} &= - \ppd{\gibbs_2}{\sigma_{\difference}} \dd{\sigma_{\difference}}{t}, \\
    \label{eq:32}
    \ddd{x_{\difference}}{t} &= - \pd{^3\gibbs_2}{\sigma_{\difference}^3} \left( \dd{\sigma_{\difference}}{t} \right)^2 - \ppd{\gibbs_2}{\sigma_{\difference}} \ddd{\sigma_{\difference}}{t}.
  \end{align}
\end{subequations}
Using~\eqref{eq:26} we see that \eqref{eq:32} can be rewritten as
\begin{equation}
  \label{eq:35}
  \ddd{x_{\difference}}{t}
  =
  -
  \pd{^3\gibbs_2}{\sigma_{\difference}^3}
  \left[
    \dd{\sigma}{t} + \ppd{\fenergy_1}{x_1} \dd{x_1}{t}
  \right]^2
  +
  \ppd{\gibbs_2}{\sigma_{\difference}}
  \left[
    \ddd{\sigma}{t} + \pd{^3\fenergy_1}{x_1^3} \left(\dd{x_1}{t}\right)^2 + \ppd{\fenergy_1}{x_1} \ddd{x_1}{t}
  \right]
\end{equation}
On the right-hand side we can---after we take all derivatives with respect to $\sigma_{\difference}$---replace all occurrences of $\sigma_{\difference}$ by
\begin{equation}
  \label{eq:37}
  \sigma_{\difference} = \sigma_1 - \sigma = - \left( \pd{\fenergy_1}{x_1} + \sigma\right),
\end{equation}
which is a consequence of~\eqref{eq:146} and the fact that the force/stress in the outer/first spring is given in terms the derivative of Helmholtz free energy, see~\eqref{eq:17}. In this sense, we can rewrite $\sigma_{\difference}$ in terms of $\sigma$ and $x_1$. On the other hand, by virtue of~\eqref{eq:18} and the evolution equation~\eqref{eq:15} we see that
\begin{equation}
  \label{eq:33}
  \ddd{x_{\difference}}{t} = \ddd{x_2}{t} - \ddd{x_1}{t} = \frac{\sigma_{\difference}}{m_2} - \ddd{x_1}{t},
\end{equation}
which together with the definition of $\sigma_{\difference}$ yields
\begin{equation}
  \label{eq:34}
  \ddd{x_{\difference}}{t} = \frac{\sigma_1 - \sigma}{m_2} - \ddd{x_1}{t}
\end{equation}
or, if rewritten in terms of the Helmholtz free energy for the first spring
\begin{equation}
  \label{eq:36}
  \ddd{x_{\difference}}{t} = - \frac{\sigma + \pd{\fenergy_1}{x_1}}{m_2} - \ddd{x_1}{t}.
\end{equation}
Now we use this expression on the left-hand side of~\eqref{eq:35}, and we arrive at
\begin{multline}
  \label{eq:38}
  - \frac{\sigma + \pd{\fenergy_1}{x_1}}{m_2} - \ddd{x_1}{t}
  =
  -
  \left.
    \pd{^3\gibbs_2}{\sigma_{\difference}^3}
  \right|_{ \sigma_{\difference} = - \left( \pd{\fenergy_1}{x_1} + \sigma\right)}
  \left[
    \dd{\sigma}{t} + \ppd{\fenergy_1}{x_1} \dd{x_1}{t}
  \right]^2
  \\
  +
  \left.
  \ppd{\gibbs_2}{\sigma_{\difference}}
  \right|_{ \sigma_{\difference} = - \left( \pd{\fenergy_1}{x_1} + \sigma\right)}
  \left[
    \ddd{\sigma}{t} + \pd{^3\fenergy_1}{x_1^3} \left(\dd{x_1}{t}\right)^2 + \ppd{\fenergy_1}{x_1} \ddd{x_1}{t}
  \right]
  .
\end{multline}
This equation includes only the variables $\sigma$, $x_1$ and their time derivatives; the variable $x_2$ explicitly characterising the behaviour of the ``internal resonator'' has been completely eliminated. Equation~\eqref{eq:38} is the sought effective constitutive relation for the nonlinear discrete mass-in-mass system---that is the nonlinear counterpart of~\eqref{eq:109}.

We note that we recover the linear case~\eqref{eq:109} discussed in Section~\ref{sec:linear} provided that we set the potentials as
\begin{subequations}
  \label{eq:21}
  \begin{align}
    \label{eq:22}
    \fenergy_1 &=_{\bydefinition} k_1 \frac{x_1^2}{2}, \\
    \label{eq:23}
    \gibbs_2 &=_{\bydefinition} \frac{1}{2k_2} \sigma_{\difference}^2.
  \end{align}
\end{subequations}
Indeed, if we do so, then~\eqref{eq:38} reduces to
\begin{equation}
  \label{eq:39}
  - \frac{\sigma + k_1 x_1}{m_2} - \ddd{x_1}{t}
  =
  \frac{1}{k_2}
  \left[
    \ddd{\sigma}{t}  + k_1 \ddd{x_1}{t}
  \right]
  ,
\end{equation}
which is the effective constitutive relation for the linear discrete mass-in-mass system, see~\eqref{eq:109}.

\subsubsection{From the energy to the effective constitutive relation}
\label{sec:from-energy-effect}
Similar to Section \ref{sec:from-energy-effect-1}, we can now reverse the reasoning regarding the interplay between the effective constitutive relation and the generalised elastic stored energy. We start with the prescribed generalised elastic stored energy, and we identify the effective constitutive relation implied by the requirement on vanishing ``entropy production''. In principle we repeat the calculations in Section~\ref{sec:from-energy-effect-1}, but in the nonlinear setting.

The generalised elastic stored energy for the fully resolved simple discrete mass-in-mass system~\eqref{eq:13} reads
\begin{equation}
  \label{eq:40}
  W
  =
  \frac{1}{2}m_2 \left( \dd{x_2}{t} \right)^2
  +
  \fenergy_1 \left( x_1\right)
  +
  \fenergy_2 \left( x_{\difference} \right)
  .
\end{equation}
In order to proceed we need to rewrite~$W$ in terms of variables $x_1$, $\dd{x_1}{t}$, $\sigma$ and $\dd{\sigma}{t}$. (We need to eliminate the variable $x_2$ related to the internal resonator.) The first step in the conversion is to rewrite~\eqref{eq:40} in terms of $\sigma_{\difference}$ and use relation~\eqref{eq:37} to replace $\sigma_{\difference}$, that is, we use the formula
\begin{equation}
  \label{eq:41}
  \sigma_{\difference}\left(x_1, \sigma\right) =  - \left( \pd{\fenergy_1}{x_1} + \sigma\right).
\end{equation}
Next, we calculate the time derivative $\dd{x_2}{t}$ using the expression~\eqref{eq:31}, that is,
\begin{equation}
  \label{eq:43}
  \dd{x_{\difference}}{t} = - \ppd{\gibbs_2}{\sigma_{\difference}} \dd{\sigma_{\difference}}{t},
\end{equation}
which in virtue of $x_{\difference} =_{\bydefinition} x_2 - x_1$, see~\eqref{eq:18},  yields
\begin{equation}
  \label{eq:42}
  \dd{x_2}{t}
  =
  \dd{x_1}{t}
  -
  \ppd{\gibbs_2}{\sigma_{\difference}} \dd{\sigma_{\difference}}{t}
  .
\end{equation}
With the usual abuse of notation, the Helmholtz free energy for the inner/second spring can be rewritten in terms of the Gibbs free energy as
\begin{equation}
  \label{eq:44}
  \fenergy_2 \left( x_{\difference} \right)
  =
  -
  \gibbs_2
  +
  \sigma_{\difference} \pd{\gibbs_2}{\sigma_{\difference}}.
\end{equation}
Using~\eqref{eq:42} and \eqref{eq:44} we can rewrite~\eqref{eq:40} as
\begin{equation}
  \label{eq:12}
  W
  =
  \frac{1}{2}
  m_2
  \left(
    \dd{x_1}{t}
    -
    \ppd{\gibbs_2}{\sigma_{\difference}} \dd{\sigma_{\difference}}{t}
  \right)^2
  +
  \fenergy_1
  +
  \left(
    -
    \gibbs_2
    +
    \sigma_{\difference} \pd{\gibbs_2}{\sigma_{\difference}}
  \right)
\end{equation}
and finally as
\begin{equation}
  \label{eq:45}
   W
   =_{\bydefinition}
      \left.
   \left(
     \frac{1}{2}
     m_2
     \left(
       \dd{}{t}
       \left(
         x_1
         -
         \pd{\gibbs_2}{\sigma_{\difference}}
       \right)
     \right)^2
     +
     \fenergy_1 
     +
     \left(
       -
       \gibbs_2
       +
       \sigma_{\difference} \pd{\gibbs_2}{\sigma_{\difference}}
     \right)
   \right)
   \right|_{\sigma_{\difference} = - \left( \pd{\fenergy_1}{x_1} + \sigma\right) }
.
 \end{equation}
 This is the generalised elastic stored energy written only in terms of $x_1$, $\dd{x_1}{t}$, $\sigma$ and $\dd{\sigma}{t}$, that is the generalised elastic stored energy of the form
 \begin{equation}
   \label{eq:115}
   W = W \left( x_1, \dd{x_1}{t}, \sigma, \dd{\sigma}{t} \right).
 \end{equation}
 Clearly, if we choose the potentials as in~\eqref{eq:21}, then the formula for the generalised elastic stored energy~\eqref{eq:45} reduces to the formula known from the linear setting~\eqref{eq:2}.

 Now we are in the position to perform a similar calculation as in Section~\ref{sec:from-energy-effect-1}. First, we observe that
 \begin{equation}
   \label{eq:47}
   \pd{}{\sigma_{\difference}}
   \left(
     -
     \gibbs_2
     +
     \sigma_{\difference} \pd{\gibbs_2}{\sigma_{\difference}}
   \right)
   =
   -
   \pd{\gibbs_2}{\sigma_{\difference}}
   +
   \pd{\gibbs_2}{\sigma_{\difference}}
   +
   \sigma_{\difference} \ppd{\gibbs_2}{\sigma_{\difference}}
   =
   \sigma_{\difference} \ppd{\gibbs_2}{\sigma_{\difference}}.
 \end{equation}
 Further, the definition~\eqref{eq:41} implies
 \begin{subequations}
   \label{eq:48}
   \begin{align}
     \label{eq:49}
     \pd{\sigma_{\difference}}{x_1} &=  - \ppd{\fenergy_1}{x_1}, \\
     \label{eq:50}
     \pd{\sigma_{\difference}}{\sigma} &= -1.
   \end{align}
 \end{subequations}
 Using the just derived formulae and the chain rule, we see that
 \begin{equation}
   \label{eq:46}
   \dd{W}{t}
   =
   m_2
   \left[
     \dd{}{t}
     \left(
       x_1
       -
       \pd{\gibbs_2}{\sigma_{\difference}}
     \right)
   \right]
   \ddd{}{t}
   \left(
     x_1
     -
     \pd{\gibbs_2}{\sigma_{\difference}}
   \right)
    +
    \pd{\psi_1}{x_1}
    \dd{x_1}{t}
    +
    \sigma_{\difference} \ppd{\gibbs_2}{\sigma_{\difference}}
    \dd{\sigma_{\difference}}{t}
    ,
  \end{equation}
  where we have kept $\sigma_{\difference}$ rather than using~\eqref{eq:41} to replace it. As in the previous section, we now exploit the stress-power formula~\eqref{eq:4}, $\dd{W}{t} = - \sigma \dd{x_1}{t}$. Combining this stress-power formula and equation \eqref{eq:46} we get
 \begin{equation}
   \label{eq:52}
   0
   =
  m_2
   \left[
     \dd{}{t}
     \left(
       x_1
       -
       \pd{\gibbs_2}{\sigma_{\difference}}
     \right)
   \right]
   \ddd{}{t}
   \left(
     x_1
     -
     \pd{\gibbs_2}{\sigma_{\difference}}
   \right)
   +
   \left(
     \sigma
     +
     \pd{\psi_1}{x_1}
   \right)
   \dd{x_1}{t}
   +
    \sigma_{\difference} \ppd{\gibbs_2}{\sigma_{\difference}}
    \dd{\sigma_{\difference}}{t}
    ,
  \end{equation}
  which can be rewritten as
  \begin{equation}
    \label{eq:53}
       0
   =
 m_2
   \left[
     \dd{}{t}
     \left(
       x_1
       -
       \pd{\gibbs_2}{\sigma_{\difference}}
     \right)
   \right]
   \ddd{}{t}
   \left(
     x_1
     -
     \pd{\gibbs_2}{\sigma_{\difference}}
   \right)
    +
    \sigma_{\difference}
    \left(
      -
      \dd{x_1}{t}
      +
      \ppd{\gibbs_2}{\sigma_{\difference}}
      \dd{\sigma_{\difference}}{t}
    \right)
  \end{equation}
  by using
  $
  -
  \sigma_{\difference}
  =
  \sigma
  +
  \pd{\psi_1}{x_1}
  $, see~\eqref{eq:41}. We rearrange the terms, and we arrive at
  \begin{equation}
    \label{eq:54}
    0
    =
    m_2
   \left[
     \dd{}{t}
     \left(
       x_1
       -
       \pd{\gibbs_2}{\sigma_{\difference}}
     \right)
   \right]
   \left[
     \ddd{}{t}
     \left(
       x_1
       -
       \pd{\gibbs_2}{\sigma_{\difference}}
     \right)
     -
      \frac{
        \sigma_{\difference}
      }{m_2}
    \right]
    .
  \end{equation}
  
  As in Section \ref{sec:from-energy-effect-1}, we assume that at any particular time, the value of $\dd{}{t}\frac{\partial g_2}{\partial \sigma_{\mathrm{diff}}}$ and $\dd{x_1}{t}$ can be arbitrarily assigned independently of each other. The rationale is essentially the same as in Section \ref{sec:from-energy-effect-1} in light of \eqref{eq:18} and \eqref{eq:19}; see the discussion after \eqref{eq:7}. Thus, the term in the second bracket of \eqref{eq:54} must be identically zero, and we end up with the effective constitutive relation in a particularly simple form
  \begin{subequations}
    \label{eq:59}
    \begin{align}
      \label{eq:60}
      \dd{}{t}
      \left(
      \dd{}{t}
      \left(
      x_1
      -
      \pd{\gibbs_2}{\sigma_{\difference}}
      \right)
      \right)
      -
      \frac{
      \sigma_{\difference}
      }{m_2}
      &=
        0, \\
      \label{eq:61}
      \sigma_{\difference} &= - \left( \pd{\fenergy_1}{x_1} + \sigma\right).
    \end{align}
  \end{subequations}
  This is the same effective constitutive relation as~\eqref{eq:38}, but now derived using the generalised elastic stored energy and the requirement that the ``entropy production'' vanishes in all mechanical processes.  Using this form we can clearly identify the origin of the individual terms in the linear effective constitutive relation~\eqref{eq:112}, see the choice of potentials in~\eqref{eq:21}. The effective constitutive relation~\eqref{eq:59} is a rate-type equation relating the displacement/strain of the outer/first mass $x_1$ and the total force/stress $\sigma$.
  
  \section{Fully nonlinear three-dimensional continuous medium}
  \label{sec:fully-three-dimens}
  Now we proceed to derive the fully three-dimensional effective constitutive relations for a \emph{continuous medium}, while the constitutive relations \emph{resemble the rate-type effective constitutive relations~\eqref{eq:59} for the one-dimensional discrete mass-in-mass system}. We go directly to the fully nonlinear setting describing finite deformations, and the linear setting (small displacement gradient theory) is subsequently obtained by linearisation.

  Our main source of inspiration for the design of a fully nonlinear three-dimensional model for a continuous medium is the approach presented in Section~\ref{sec:non-linear}. In particular, we are motivated by the structure of the generalised elastic stored energy~\eqref{eq:45}, which reads
  \begin{equation}
    \label{eq:51}
    W
   =
   \left.
   \left(
     \frac{1}{2}
     m_2
     \left(
       \dd{}{t}
       \left(
         x_1
         -
         \pd{\gibbs_2}{\sigma_{\difference}}
       \right)
     \right)^2
     +
     \fenergy_1 
     +
     \left(
       -
       \gibbs_2
       +
       \sigma_{\difference} \pd{\gibbs_2}{\sigma_{\difference}}
     \right)
   \right)
   \right|_{\sigma_{\difference} = - \left( \pd{\fenergy_1}{x_1} + \sigma\right) },
 \end{equation}
 and we want to obtain a rate-type equation that mimics~\eqref{eq:59}, that is
 \begin{subequations}
   \label{eq:148}
    \begin{align}
      \label{eq:149}
      \dd{}{t}
      \left(
      \dd{}{t}
      \left(
      x_1
      -
      \pd{\gibbs_2}{\sigma_{\difference}}
      \right)
      \right)
      -
      \frac{
      \sigma_{\difference}
      }{m_2}
      &=
        0, \\
      \label{eq:150}
      \sigma_{\difference} &= - \left( \pd{\fenergy_1}{x_1} + \sigma\right).
    \end{align}
  \end{subequations}
We point out that our approach is not based on a \emph{homogenisation} and \emph{continualisation} of a simple discrete mass-in-mass system. Instead, we are \emph{loosely} motivated by the key qualitative feature behind~\eqref{eq:148}, namely by the existence of the generalised elastic stored energy with a special structure, and we make phenomenological assumptions without a detailed discussion of a particular arrangement of ``voids'' and ``internal resonators'' (inner/outer masses) and ``springs'' in the given fully resolved continuous medium.

The lack of exact one-to-one correspondence between the proposed model and a fully resolved microscopic model will already be seen at the level of the density. The density of the continuous medium will be the overall (apparent) density $\rho$, that is the density that enters the body force term in the gravitational field and the momentum. We do not address the question of how the density of the medium is related to the partial densities $\rhoA$ and $\rhoB$ of the outer and inner ``masses'' and their distribution in an elementary cell, see Figure~\eqref{fig:internal-arrangement}. This question would require a detailed analysis of the voids arrangement and so forth, which is outside the scope of the current work. In this sense, the model proposed in this study is a purely phenomenological one.

Nevertheless, since the phenomenological model respects the key qualitative features behind the simple discrete mass-in-mass systems, it inherits the peculiar behaviour of simple mass-in-mass systems with respect to the wave propagation.

\subsection{Initial remarks}
\label{sec:initial-remarks}

We derive the model in the Eulerian setting---we work with quantities available in the current configuration only. This helps us to reduce detailed microscopic kinematic analysis of metamaterial behaviour. In order to proceed with a rate-type model for finite deformations in the Eulerian setting, we need to pick an objective tensorial rate and a convenient strain measure. In this regard, we choose the \emph{logarithmic corotational derivative} and the \emph{Hencky strain}. This choice is motivated by the following considerations.

The logarithmic corotational derivative is given by the formula
  \begin{equation}
    \label{eq:62}
    \logfid{\overline{\generictensor}}
    =_{\bydefinition}
    \dd{\generictensor}{t}
    +
    \generictensor
    \logspin
    -
    \logspin
    \generictensor
    ,
  \end{equation}
  where $\logspin$ denotes the logarithmic spin (a skew-symmetric tensor), given by the formula
  \begin{equation}
    \label{eq:logarithmic-spin-definition}
    \logspin
    =_{\bydefinition}
    \gradasym
    +
    \sum_{\substack{\sigma, \tau =1 \\ \sigma \not = \tau}}^3
    \left[
      \left(
        \frac{1 + \frac{b_\sigma}{b_\tau}}{1 + \frac{b_\sigma}{b_\tau}}
        +
        \frac{2}{\ln \frac{b_\sigma}{b_\tau}}
      \right)
      \tensorq{P}_\sigma
      \gradsym
      \tensorq{P}_\tau
    \right]
    ,
  \end{equation}
  where $\gradsym$ and $\gradasym$ denote the symmetric/skew-symmetric part of the Eulerian velocity gradient~$\gradv$, $b_{\sigma}$ denotes the $\sigma$-th eigenvalue of the left Cauchy--Green tensor $\lcg = \fgrad \transpose{\fgrad}$ and $\tensorq{P}_\sigma$  denotes the projection to the corresponding eigenspace, see~\cite[Eq. 41]{xiao.h.bruhns.ot.ea:logarithmic} for in-depth discussion. The key property of the logarithmic corotational derivative is that \emph{the logarithmic corotational derivative of the Hencky strain tensor~$\henckystrain$},
  \begin{equation}
    \label{eq:64}
    \henckystrain
    =_\bydefinition
    \frac{1}{2}
    \ln \lcg
    ,
  \end{equation}
  \emph{yields directly the symmetric part of the velocity gradient~$\gradsym$}, that is we have the identity
  \begin{equation}
    \label{eq:63}
    \logfid{\overline{\henckystrain}}
    =
    \gradsym.
  \end{equation}
This property makes the logarithmic derivative suitable for rate-type formulations of elasticity, see, for example, \cite{xiao.h.bruhns.ot.ea:logarithmic} and \cite{xiao.h.bruhns.ot.ea:natural}. In fact, the Hencky strain is the unique choice amongst Eulerian Hill strains for which an objective corotational derivative can be found that yields the stretching~$\gradsym$, see \cite{xiao.h.bruhns.ot.ea:logarithmic}. For additional discussion of objective derivatives and especially the logarithmic corotational derivative, see~\cite{xiao.h.bruhns.ot.ea:logarithmic}, \cite{bruhns.ot.meyers.a.ea:on}, \cite{kolev.b.desmorat.r:objective} and~\cite{bathory.m.bulcek.ea:new}. Furthermore, the Hencky strain is a natural choice for the \emph{dual variable} in the Legendre transform (Helmholtz free energy versus Gibbs free energy), see \cite{fitzgerald.je:tensorial}, \cite{hoger.a:stress}, \cite{haupt.p.tsakmakis.c:on}, \cite{xiao.h.bruhns.ot.ea:explicit}, \cite{gokulnath.c.saravanan.u.ea:representations} and~\cite{prusa.v.rajagopal.kr.ea:gibbs} to name a few.

Concerning continuum thermodynamics we recall the following facts; for details see \cite{malek.j.prusa.v:derivation}, \cite{truesdell.c.noll.w:non-linear*1} or any standard work on continuum thermodynamics. The standard evolution equation for the internal energy density $\ienergy = \ienergy \left(\entropy, y_1, \dots, y_n \right)$ reads
  \begin{equation}
    \label{eq:24}
    \rho \dd{\ienergy}{t}
    =
    \tensordot{\cstress}{\gradsym}
    -
    \divergence \hfluxc
    ,
  \end{equation}
  where $\hfluxc$ denotes the heat flux, $\entropy$ denotes the entropy density and $\left\{ y_i \right\}_{i=1}^n$ denote other quantities the internal energy density can depend on. We work with internal energy density \emph{per unit mass}, that is $[\ienergy] = \unitfrac{J}{kg}$, and the internal energy of continuous medium occupying the volume $V$ is obtained as $\int_V \rho \ienergy \, \cvolumee$. The same holds for other thermodynamic potentials. Using the standard definition of the thermodynamic temperature, see for example~\cite{callen.hb:thermodynamics},
  \begin{equation}
    \label{eq:105}
    \temp =_{\bydefinition} \pd{\ienergy}{\entropy},
  \end{equation}
  and using the chain rule on the left-hand side of~\eqref{eq:24}, we get the evolution equation for the entropy $\entropy$ in the form
  \begin{equation}
    \label{eq:147}
    \rho \dd{\entropy}{t}
    +
    \divergence
    \left(
      \frac{\hfluxc}{\temp}
    \right)
    =
    \frac{1}{\temp}
    \left(
      \tensordot{\cstress}{\gradsym}
      -
      \rho
      \pd{\ienergy}{y_i} \dd{y_i}{t}
    \right)
    -
    \frac{
      \vectordot{\hfluxc}{\nabla \temp}
    }
    {
      \temp^2
    }
    .
  \end{equation}
  Replacing the internal energy $\ienergy$ by the Helmholtz free energy $\fenergy$ introduced via the Legendre transform,
  \begin{equation}
    \label{eq:66}
    \fenergy(\temp, y_1, \dots, y_n) =_{\bydefinition} \ienergy(\entropy(\temp, y_1, \dots, y_n), y_1, \dots, y_n) - \temp \entropy(\temp, y_1, \dots, y_n),
  \end{equation}
  we finally get the entropy evolution equation in the form
  \begin{equation}
    \label{eq:65}
       \rho \dd{\entropy}{t}
    +
    \divergence
    \left(
      \frac{\hfluxc}{\temp}
    \right)
    =
    \frac{1}{\temp}
    \underbrace{
      \left(
        \tensordot{\cstress}{\gradsym}
        -
        \rho
        \pd{\fenergy}{y_i} \dd{y_i}{t}
      \right)
    }_{\entprodctemp_{\text{mech}}}
    -
    \frac{
      \vectordot{\hfluxc}{\nabla \temp}
    }
    {
      \temp^2
    }
    .
  \end{equation}
  Note that we now use the sign convention and Legendre transform definition common in continuum thermodynamics. 
  
  An \textit{elastic solid} is a solid that does not produce entropy in purely mechanical processes, which means that the corresponding entropy production term $\entprodctemp_{\text{mech}}$ on the right-hand side of~\eqref{eq:65} is always zero,
  \begin{equation}
    \label{eq:67}
    \tensordot{\cstress}{\gradsym}
    -
    \rho
    \pd{\fenergy}{y_i} \dd{y_i}{t}
    =
    0
    .
  \end{equation}
  Equation~\eqref{eq:67} allows us to derive the constitutive relation from the formula for the Helmholtz free energy/Gibbs free energy, and it constitutes the continuum mechanics counterpart to the stress-power equation~\eqref{eq:4}.

  \subsection{From the energy to the effective constitutive relation}
  \label{sec:from-energy-effect-2}
  Concerning a single elastic continuous medium, the following relations hold in the standard case of isotropic hyperelastic/Green elastic solid
  \begin{subequations}
    \label{eq:81}
    \begin{align}
      \label{eq:82}
      \thcstressrho &= \pd{\fenergy}{\henckystrain}, \\
      \label{eq:83}
      \gibbs &= \fenergy - \tensordot{\thcstressrho}{\henckystrain}, \\
      \label{eq:84}
      - \henckystrain &= \pd{\gibbs}{\thcstressrho},
    \end{align}
  \end{subequations}
  see~\cite{prusa.v.rajagopal.kr.ea:gibbs}, where the symbol $\tensordot{\generictensor}{\tensorq{B}} =_{\bydefinition} \Tr \left( \generictensor \transpose{\tensorq{B}} \right)$ denotes the standard dot product on the space of second order tensors. The remaining notation in~\eqref{eq:81} follows~\cite{prusa.v.rajagopal.kr.ea:gibbs}, in particular the symbol~$\thcstressrho$ denotes the Cauchy stress tensor $\thcstress$ divided by the density $\rho$ in the current configuration,
\begin{equation}
  \label{eq:3}
  \thcstressrho =_{\bydefinition} \frac{\thcstress}{\rho}.
\end{equation}
  The subscript $(\cdot)_\mathrm{th}$ denotes the thermodynamically constituted part of the stress, that is the part of the stress that comes from the derivative of a thermodynamic (energy-like) potential. The first formula~\eqref{eq:82} means that the stress is derivable via the differentiation of the Helmholtz free energy expressed as a function of the Hencky strain tensor, and the last formula~\eqref{eq:84} is the dual statement of the same in terms of the dual potential---the Gibbs free energy. Formula~\eqref{eq:83} is the Legendre transformation between the Helmholtz free energy and the Gibbs free energy, the compact formula for the Legendre transform is the consequence of the choice of dual variables $\thcstressrho$ and $\henckystrain$.

  Concerning the metamaterial model, we mimic the approach used in Section~\ref{sec:non-linear}. The first continuum (loosely identified with the outer mass $m_1$ and spring $k_1$ in the simple discrete mass-in-mass model) is characterised via its Helmholtz free energy $\fenergyA \left(\henckystrainA, \cdot\right)$, and we identify \emph{the Hencky strain for the first continuum~$\henckystrainA$ with the overall Hencky strain $\henckystrain$},
  \begin{equation}
    \label{eq:55}
    \henckystrain \equiv \henckystrainA.
  \end{equation}
  This corresponds to the transition from the fully resolved system, Figure~\eqref{fig:mass-in-mass-a}, to its effective counterpart, Figure~\eqref{fig:mass-in-mass-c}. The particular internal arrangement of the inner mass (internal resonator) is not seen on the macroscopic level, we are just tracking the motion of the outer shell, the presence of the inner mass manifests itself only in terms of forces/stresses acting on the shell.  Assuming that the outer/first continuum is a homogeneous isotropic hyperelastic/Green elastic solid described by the Helmholtz free energy, we specialise~\eqref{eq:82} for the outer/first continuum as
  \begin{equation}
    \label{eq:56}
    \thcstressrhoA = \pd{\fenergyA}{\henckystrainA},
  \end{equation}
  Isotropy implies that $\fenergyA$ is a scalar function of invariants of $\henckystrainA$. The notation is clear, $\thcstressrhoA =_{\bydefinition} \frac{\thcstressA}{\rhoA}$, see~\eqref{eq:3}, where~$\thcstressA$ is the outer/first continuum contribution to the overall stress, and where $\rhoA$ is the density of the outer/first continuum. We identify $\rhoA$ with the overall density $\rho$ of the material of interest, that is we set
  \begin{equation}
    \label{eq:57}
    \rhoA \equiv \rho.
  \end{equation}
  This assumption is consistent with~\eqref{eq:55}, and later we see that it implies that if the ``internal resonators'' are absent, then the material of interest behaves as an ordinary hyperelastic/Green elastic solid with the Helmholtz free energy $\fenergyA$.

  The second continuum (the inner mass, the internal resonator; loosely identified with the inner mass~$m_2$ and the spring~$k_2$ in the simple discrete mass-in-mass model) is again assumed to be a homogeneous isotropic hyperelastic/Green elastic solid. Unlike in the previous case we however characterise the solid via its Gibbs free energy $\gibbsB \left( \thcstressrhoDiff, \cdot \right)$ specified in terms of the corresponding stress~$\thcstressrhoDiff$, that is we are not using the standard characterisation by the means of Helmholtz free energy. To some extent, the use of Gibbs free energy for the inner/second continuum allows us to abstain from detailed description of kinematics of the inner/second continuum---see the discussion in the discrete case following equation~\eqref{eq:29} in Section~\ref{sec:from-fully-resolved}. Following the dual potential relation~\eqref{eq:84} specialised for the inner/second continuum we have
  \begin{subequations}
    \label{eq:86}
    \begin{align}
      \label{eq:87}
      -\henckystrainDiff &= \pd{\gibbsB}{\thcstressrhoDiff}, \\
      \label{eq:88}
      \henckystrainDiff &= \henckystrainA - \henckystrainB, \\
      \label{eq:89}
      \fenergyB &= \gibbsB + \tensordot{\thcstressrhoDiff}{\henckystrainDiff}.
    \end{align}
  \end{subequations}
 Isotropy of the inner/second continuum implies that $\gibbsB$ is a scalar function of invariants of $\thcstressrhoDiff$. The notation is clear, $\thcstressrhoDiff =_{\bydefinition} \frac{\thcstressDiff}{\rhoB}$, see~\eqref{eq:3}, where $\thcstressDiff$ is the inner/second continuum contribution to the overall stress, and where~$\rhoB$ is the density of the inner/second continuum. Formulae~\eqref{eq:87} and \eqref{eq:88} are clearly motivated by the formulae~\eqref{eq:18} and \eqref{eq:19}. (The sign change of \eqref{eq:88} compared to \eqref{eq:18} is due to a different sign convention in the stress definition, compare~\eqref{eq:17} and~\eqref{eq:82}.) In particular, the formula~\eqref{eq:88} is \emph{yet another constitutive assumption}, and this constitutive assumption is in line with the approach taken in Section~\ref{sec:non-linear}---note that for commuting tensors~$\henckystrainA$ and~$\henckystrainB$ the tensor exponential transforms the ``sum'' to a ``composition''. The constitutive assumption~\eqref{eq:88} can be seen as a characterisation of the kinematics of the internal resonator.

Now we need to analyse the decomposition/split of overall stress $\thcstress$ to the stress induced by the outer/first continuum~$\thcstressA$ and the inner/second continuum $\thcstressDiff$. In other words, we need an analogue of stress decomposition~\eqref{eq:146} used in the case of discrete mass-in-mass system. This would in principle require us to carefully analyse the particular interaction between the internal resonator and the surrounding medium, which we want to avoid. We replace this analysis by yet another constitutive assumption, namely with the assumption
  \begin{equation}
    \label{eq:58}
    \thcstressrho = \thcstressrhoA + \thcstressrhoDiff.
  \end{equation}
  This assumption is in line with~\eqref{eq:146}, the sign difference between~\eqref{eq:58} and~\eqref{eq:146}  with respect to~\eqref{eq:146} is again due to a different sign convention. In essence, \eqref{eq:58} means that the overall density rescaled Cauchy stress tensor~$\thcstressrho$---an unknown quantity for which we need a constitutive relation---is given by a particular density weighted sum of the stress contributions by the inner/second and the outer/first continuous medium.

  Exploiting the existence of the potential, see~\eqref{eq:56}, we see that~\eqref{eq:58} yields
  \begin{equation}
    \label{eq:90}
    \thcstressrhoDiff = \thcstressrho - \pd{\fenergyA}{\henckystrainA}.
  \end{equation}
  Furthermore, using \eqref{eq:87} and \eqref{eq:88} we get
  \begin{equation}
    \label{eq:91}
    \henckystrainB = \henckystrainA + \pd{\gibbs}{\thcstressrhoDiff}.
  \end{equation}
  
  Now we are in a position to proceed with the specification of a \emph{generalised} Helmholtz free energy for the whole continuum. (As we are interested in purely mechanical response, we deal only with the purely mechanical part of the Helmholtz free energy. If necessary a standard purely thermal part of Helmholtz free energy can be added.) Motivated by the generalised energy formula in the discrete mass-in-mass system in Section~\ref{sec:non-linear}---see formula~\eqref{eq:45}---we start with the generalised Helmholtz free energy in the form
  \begin{multline}
    \label{eq:68}
    \fenergy
    \left(
      \henckystrainA,
      \thcstressrho
    \right)
    =_{\bydefinition}
    \frac{1}{2}
    \tensordot{
      \logfid{\overline{\left( \henckystrainA + \left. \pd{\gibbsB}{\thcstressrhoDiff} \right|_{\thcstressrhoDiff = \thcstressrho - \pd{\fenergyA}{\henckystrainA}} \right)}}
    }
    {
      \ratestiffnesstensor
      \logfid{\overline{\left( \henckystrainA + \left. \pd{\gibbsB}{\thcstressrhoDiff} \right|_{\thcstressrhoDiff = \thcstressrho - \pd{\fenergyA}{\henckystrainA}} \right)}}
    }
    \\
    +
    \fenergyA
    +
    \left.
    \left(
      \gibbsB
      -
      \tensordot{
        \thcstressrhoDiff
      }
      {
        \pd{\gibbsB}{\thcstressrhoDiff}
      }
    \right)
    \right|_{\thcstressrhoDiff = \thcstressrho - \pd{\fenergyA}{\henckystrainA}},
  \end{multline}
  where $\ratestiffnesstensor$ is a constant symmetric positive definite fourth-order ``rate stiffness'' tensor. Here, positive definiteness means that $\tensordot{\generictensor}{\ratestiffnesstensor \generictensor} > 0$ holds for all non-zero second order tensors $\generictensor$, while symmetric means that $\tensordot{\generictensor_1}{\ratestiffnesstensor \generictensor_2} = \tensordot{\generictensor_2}{\ratestiffnesstensor \generictensor_1}$ holds for all non-zero second order tensors $\generictensor_1$ and $\generictensor_2$. Recall that in virtue of Legendre transform we have
  \begin{equation}
    \label{eq:93}
    \fenergyB
    =
    \gibbsB
    -
    \tensordot{
      \thcstressrhoDiff
    }
    {
      \pd{\gibbsB}{\thcstressrhoDiff}
    }
    ,
  \end{equation}
  and this is the rationale behind the last term in~\eqref{eq:68}. The first term in~\eqref{eq:68} is designed as an analogue to the first term in~\eqref{eq:45}, but we have taken the liberty to use a generic quadratic form induced by $\ratestiffnesstensor$ instead of mere (tensorial) dot product. Furthermore in the energy formula~\eqref{eq:68} we use sign conventions for the potentials common in continuum mechanics---this is the reason for some sign changes with respect to the energy formula~\eqref{eq:45} for the mass-in-mass system. 

  Having identified the \emph{ansatz} for the generalised Helmholtz free energy, we proceed with the thermodynamically based derivation of the corresponding rate-type constitutive relation. We divide~\eqref{eq:67} by $\rho$, we interpret $\henckystrainA$ as the full Hencky strain $\henckystrain$ for the material, see~\eqref{eq:55}, and we use identity~\eqref{eq:63} for the logarithmic corotational derivative. This yields
  \begin{equation}
    \label{eq:73}
    \tensordot{\cstressrho}{\logfid{\overline{\henckystrainA}}}
    -
    \pd{\fenergy}{y_i} \dd{y_i}{t}
    =
    0
    .
  \end{equation}
  Here $\cstressrho$ denotes the overall Cauchy stress in the material divided by the material density $\rho$,
  \begin{equation}
    \label{eq:85}
    \cstressrho =_{\bydefinition} \frac{\cstress}{\rho}.
  \end{equation}
  Note that since the overall Cauchy stress $\cstress$ can have a contribution not linked to a thermodynamic energy-like potential---this happens for example in viscoelastic materials---we can not \emph{a priori} identify $\cstressrho = \thcstressrho$. This comes later as yet another constitutive assumption. In~\eqref{eq:73} we further use the chain rule and the specific \emph{ansatz} for the Helmholtz free energy~\eqref{eq:68}. This yields
  \begin{multline}
    \label{eq:74}
    \tensordot{\cstressrho}{\logfid{\overline{\henckystrainA}}}
    -
    \pd{\fenergy}{y_i} \dd{y_i}{t}
    =
    \tensordot{\cstressrho}{\logfid{\overline{\henckystrainA}}}
    -
    \tensordot{
      \logfid{\overline{\left( \henckystrainA + \pd{\gibbsB}{\thcstressrhoDiff} \right)}}
    }
    {
      \ratestiffnesstensor
      \dd{}{t}
      \left(
        \logfid{\overline{\left( \henckystrainA + \pd{\gibbsB}{\thcstressrhoDiff} \right)}}
      \right)
    }
    \\
    -
    \tensordot{
      \pd{\fenergyA}{\henckystrainA}
    }
    {
      \dd{\henckystrainA}{t}
    }
    -
    \tensordot{
      \underbrace{
        \left(
          -
          \tensordot{
            \thcstressrhoDiff
          }
          {
            \ppd{\gibbsB}{\thcstressrhoDiff}
          }
        \right)
      }_{
        \pd{\fenergyB}{\thcstressrhoDiff}
      }
    }
    {
      \dd{\thcstressrhoDiff}{t}
    }
    .
  \end{multline}
  Here we are exploiting an analogue to~\eqref{eq:47} which allows us to simplify the formula for the derivative of the last term in the generalised Helmholtz free energy~\eqref{eq:68}.

  We make the simplifying assumption that the fourth-order tensor $\ratestiffnesstensor$ represents a \emph{linear isotropic tensor valued function}, that is, $\ratestiffnesstensor \generictensor = a \generictensor + b \left(\Tr \generictensor \right) \identity$ holds for all $\generictensor$. (Scalar quantities $a$ and $b$ are some constants.) This implies that $\ratestiffnesstensor \generictensor$ commutes with $\generictensor$, and thanks to the properties of the logarithmic corotational derivative we can rewrite the last formula as\footnote{We use the fact that for the logarithmic corotational derivative, the isotropic fourth order tensor $\ratestiffnesstensor$ and a symmetric tensor $\generictensor$ we have
    \begin{equation}
      \label{eq:71}
      \tensordot{\left(\ratestiffnesstensor \generictensor \right)}{\dd{\generictensor}{t}}
      =
      \tensordot{\left(\ratestiffnesstensor \generictensor \right)}
      {
        \left(
          \dd{\generictensor}{t}
          +
          \generictensor
          \logspin
          -
          \logspin
          \generictensor
        \right)
      }
      =
      \tensordot{\left(\ratestiffnesstensor \generictensor \right)}{\logfid{\overline{\generictensor}}}
      .
    \end{equation}
    The equality holds because the product $\tensordot{\left(\ratestiffnesstensor \generictensor \right)}{\left( \generictensor \logspin \right)}$ reduces to $\tensordot{\left( a \generictensor^2 + b \left(\Tr \generictensor \right) \generictensor \right)}{\logspin}$, which is the product of a symmetric tensor and a skew-symmetric spin tensor $\logspin$. This in fact holds for any corotational derivative due to the skew-symmetry of the spin tensor. Similar reasoning also applies in~\eqref{eq:74}---since $\psi_1$ and $g_2$ are isotropic functions we can replace the material time derivatives $\dd{\henckystrain_1}{t}$ and $\dd{\thcstressrhoDiff}{t}$ by the logarithmic corotational derivatives in \eqref{eq:74}.  
  }
  \begin{equation}
    \label{eq:75}
    0
    =
    \tensordot{\cstressrho}{\logfid{\overline{\henckystrainA}}}
    -
    \tensordot{
      \logfid{\overline{\left( \henckystrainA + \pd{\gibbsB}{\thcstressrhoDiff} \right)}}
    }
    {
      \ratestiffnesstensor
      \logfidd{\overline{\left( \henckystrainA + \pd{\gibbsB}{\thcstressrhoDiff} \right)}}
    }
    -
    \tensordot{
      \pd{\fenergyA}{\henckystrainA}
    }
    {
      \logfid{\overline{\henckystrainA}}
    }
    +
    \tensordot{
      \left(
        \tensordot{
          \thcstressrhoDiff
        }
        {
          \ppd{\gibbsB}{\thcstressrhoDiff}
        }
      \right)
    }
    {
      \logfid{\overline{\thcstressrhoDiff}}
    }
  \end{equation}
  We rearrange the terms, and we get
  \begin{equation}
    \label{eq:76}
    0
    =
    -
    \tensordot{
      \logfid{\overline{\left( \henckystrainA + \pd{\gibbsB}{\thcstressrhoDiff} \right)}}
    }
    {
      \ratestiffnesstensor
      \logfidd{\overline{\left( \henckystrainA + \pd{\gibbsB}{\thcstressrhoDiff} \right)}}
    }
    -
    \tensordot{
      \left(
        \pd{\fenergyA}{\henckystrainA}
        -
        \cstressrho
      \right)
    }
    {
      \logfid{\overline{\henckystrainA}}
    }
    +
    \tensordot{
      \thcstressrhoDiff
    }
    {
      \left(
        \tensordot{
          \ppd{\gibbsB}{\thcstressrhoDiff}
        }
        {
          \logfid{\overline{\thcstressrhoDiff}}
        }
      \right)
    }
    .
  \end{equation}

  At the moment the key term is the second term
  $
  \tensordot{
    \left(
      \pd{\fenergyA}{\henckystrainA}
      -
      \cstressrho
    \right)
  }
  {
    \logfid{\overline{\henckystrainA}}
  }$. We recall that $\cstressrho$ denotes the rescaled Cauchy stress tensor characterising the overall response of the whole continuum---the quantity for which we want to find a constitutive relation. In particular, we want to find a relation between this stress tensor and the generalised Helmholtz free energy~\eqref{eq:68}. We first assume that the overall stress $\cstressrho$ is given as a composition of a thermodynamic part $\thcstressrho$ related to the derivative of the generalised Helmholtz free energy and a dissipative part $\discstressrho$,
  \begin{equation}
    \label{eq:161}
    \cstressrho = \thcstressrho + \discstressrho.
  \end{equation}
  If we do so, the critical term would read
  \begin{equation}
    \label{eq:162}
    -
    \tensordot{
      \left(
        \pd{\fenergyA}{\henckystrainA}
        -
        \cstressrho
      \right)
    }
    {
      \logfid{\overline{\henckystrainA}}
    }
    =
    -
    \tensordot{
    \left(
      \pd{\fenergyA}{\henckystrainA}
      -
      \thcstressrho
    \right)
  }
  {
    \logfid{\overline{\henckystrainA}}
  }
  +
  \tensordot{\discstressrho}{ \logfid{\overline{\henckystrainA}}}.
\end{equation}
Choosing for example $\discstressrho =_{\bydefinition} 2 \mu \logfid{\overline{\henckystrainA}}$ with $\mu > 0$ would in virtue of~\eqref{eq:55} and~\eqref{eq:63} yield
\begin{equation}
  \label{eq:163}
  -
     \tensordot{
      \left(
        \pd{\fenergyA}{\henckystrainA}
        -
        \cstressrho
      \right)
    }
    {
      \logfid{\overline{\henckystrainA}}
    }
    =
    -
     \tensordot{
    \left(
      \pd{\fenergyA}{\henckystrainA}
      -
      \thcstressrho
    \right)
  }
  {
    \logfid{\overline{\henckystrainA}}
  }
  +
  2 \mu
  \tensordot{\gradsym}{\gradsym},
\end{equation}
where the last term in non-negative. This would lead to a non-trivial entropy production term and subsequently to a model for a viscoelastic solid material. However, we are interested in deriving constitutive relations for a material that does not produce entropy in mechanical processes. We thus assume that the rescaled Cauchy stress tensor $\cstressrho$ is \emph{identical to the thermodynamically derived rescaled Cauchy stress tensor}~$\thcstressrho$, that is instead of \eqref{eq:161} we simply assume
  \begin{equation}
    \label{eq:72}
    \cstressrho \equiv \thcstressrho.
  \end{equation}
  Using this assumption we see that the second term on the right-hand side of~\eqref{eq:76} can be rewritten as
  \begin{equation}
    \label{eq:77}
     -
    \tensordot{
      \left(
        \pd{\fenergyA}{\henckystrainA}
        -
        \cstressrho
      \right)
    }
    {
      \logfid{\overline{\henckystrainA}}
    }
    =
    \tensordot{
      \underbrace{
        \left(
          -
          \pd{\fenergyA}{\henckystrainA}
          +
          \thcstressrho
        \right)
      }_{\thcstressrhoDiff}
    }
    {
      \logfid{\overline{\henckystrainA}}
    }
    =
    \tensordot{
      \thcstressrhoDiff
    }
    {
      \logfid{\overline{\henckystrainA}}
    }
    ,
  \end{equation}
  where we have exploited the stress decomposition assumption, see~\eqref{eq:58} and~\eqref{eq:90}. Formula~\eqref{eq:77} then allows us to rewrite~\eqref{eq:76} as
  \begin{equation}
    \label{eq:78}
       0
    =
    -
    \tensordot{
      \logfid{\overline{\left( \henckystrainA + \pd{\gibbsB}{\thcstressrhoDiff} \right)}}
    }
    {
      \ratestiffnesstensor
      \logfidd{\overline{\left( \henckystrainA + \pd{\gibbsB}{\thcstressrhoDiff} \right)}}
    }
    +
    \tensordot{
      \thcstressrhoDiff
    }
    {
      \left(
        \logfid{\overline{\henckystrainA}}
        +
        \tensordot{
          \ppd{\gibbsB}{\thcstressrhoDiff}
        }
        {
          \logfid{\overline{\thcstressrhoDiff}}
        }
      \right)
    }.
  \end{equation}
  Exploiting one again the properties of corotational derivatives and the isotropy assumptions, we finally end up with the equation
  \begin{equation}
    \label{eq:79}
    0
    =
    \tensordot{
      \logfid{\overline{\left( \henckystrainA + \pd{\gibbsB}{\thcstressrhoDiff} \right)}}
    }
    {
      \left(
        -
        \ratestiffnesstensor
        \logfidd{\overline{\left( \henckystrainA + \pd{\gibbsB}{\thcstressrhoDiff} \right)}}
        +
        \thcstressrhoDiff
      \right)
    }
    ,
  \end{equation}
  wherein the right-hand side gives us an explicit formula for the entropy production $\entprodctemp_{\text{mech}}$ in mechanical processes.
  
  As in Section~\ref{sec:from-energy-effect}, see equation \eqref{eq:54}, we must choose the constitutive relation in such a way that~\eqref{eq:79} holds, guaranteeing zero entropy production in all mechanical processes.  We assume that at any particular time, the values of 
  \begin{align}
    \label{eq:155}
    \logfid{\overline{\henckystrainA}}
    \quad
    \text{and}
    \quad 
    \logfid{\overline{\pd{\gibbsB}{\thcstressrhoDiff}}}
  \end{align}
 can be arbitrarily assigned, independently of each other. The rationale is essentially the same as in Section \ref{sec:from-energy-effect-1} and Section \ref{sec:from-energy-effect} in light of \eqref{eq:91}; see the discussion after \eqref{eq:7}. Thus, the second term appearing in \eqref{eq:79} must vanish: 
  \begin{equation}
    \label{eq:80}
    -
    \ratestiffnesstensor
    \logfidd{
      \overline{
        \left(
          \henckystrainA
          +
          \left.
            \pd{\gibbsB}{\thcstressrhoDiff}
          \right|_{\thcstressrhoDiff = \thcstressrho - \pd{\fenergyA}{\henckystrainA}}
        \right)
      }
    }
    +
    \left.
      \thcstressrhoDiff
    \right|_{\thcstressrhoDiff = \thcstressrho - \pd{\fenergyA}{\henckystrainA}}
    =
    \tensorzero.
  \end{equation}
  We recall that we identify $\henckystrainA \equiv \henckystrain$, where $\henckystrain$ is the overall Hencky strain, and thus, equation~\eqref{eq:80} is the sought rate-type constitutive relation and a tensorial counterpart of~\eqref{eq:59}. Indeed, the rate-type equation~\eqref{eq:80} gives us a relation between the Hencky strain $\henckystrain$ and the Cauchy stress tensor for the whole continuum $\cstress$. Recall that we have~\eqref{eq:85} and~\eqref{eq:72}. Furthermore, we note that the term twice differentiated by the logarithmic corotational derivative is in fact~$\henckystrainB$ rewritten in terms of the primitive quantities
  \begin{equation}
    \label{eq:133}
    \henckystrainB
    =
    \henckystrainA
    +
    \left.
      \pd{\gibbsB}{\thcstressrhoDiff}
    \right|_{\thcstressrhoDiff = \thcstressrho - \pd{\fenergyA}{\henckystrainA}},
  \end{equation}
  see~\eqref{eq:91}.

\subsection{Summary of the effective constitutive relations}
\label{sec:effect-const-relat}
With the identification~\eqref{eq:55}---the overall Hencky strain is the same as the Hencky strain $\henckystrain$ for the outer/first continuum---and~\eqref{eq:72}---the overall rescaled Cauchy stress tensor $\cstressrho$ for the whole material originates from the thermodynamic potential only---we can rewrite~\eqref{eq:80} in the final form as
\begin{subequations}
  \label{eq:156}
  \begin{align}
    \label{eq:157}
    -
    \ratestiffnesstensor
    \logfidd{
      \overline{
        \left(
          \henckystrain
          +
    \pd{\gibbsB}{\thcstressrhoDiff}
        \right)
      }
    }
    +
    \thcstressrhoDiff
    &=
    \tensorzero
      ,
      \\
    \label{eq:158}
    \thcstressrhoDiff &= \cstressrho - \pd{\fenergyA}{\henckystrain}.
  \end{align}
\end{subequations}

The constitutive relation \eqref{eq:156} is a generalisation of~\eqref{eq:59} for a fully three-dimensional nonlinear (finite deformations) continuous medium. This rate-type constitutive relation describes the evolution of the overall Cauchy stress tensor $\cstressrho \equiv \thcstressrho$ and the overall Hencky strain tensor $\henckystrain$; recall that $\cstressrho = \frac{\cstress}{\rho}$. No explicit reference to a microstructure is present in~\eqref{eq:156}, and we can in principle ignore all microstructural considerations that initially motivated~\eqref{eq:156}. In this sense~\eqref{eq:156} represents a self-contained rate-type constitutive relation for an elastic solid---it is derived from the generalized Helmholtz free energy~\eqref{eq:68} and it describes a continuous medium that does not produce entropy in mechanical processes.

Naturally, the rate-type constitutive relation~\eqref{eq:156} must be solved together with the (Eulerian) balance of linear momentum and the evolution equation for the Hencky strain tensor~\eqref{eq:63}; the full system of evolution equations thus reads
\begin{subequations}
  \label{eq:evolution-equations-system}
  \begin{align}
    \label{eq:169}
    \dd{\rho}{t} + \rho \divergence \vec{v}
    &=0,
      \\
    \label{eq:164}
    \rho \dd{\vec{v}}{t}
    &=
      \divergence \cstress + \rho \vec{b},
    \\
    \label{eq:167}
    \logfid{\overline{\henckystrain}}
    &=
      \gradsym,
    \\
    \label{eq:70}
    -
    \ratestiffnesstensor
    \logfidd{
      \overline{
        \left(
          \henckystrain
          +
    \pd{\gibbsB}{\thcstressrhoDiff}
        \right)
      }
    }
    +
    \thcstressrhoDiff
    &=
    \tensorzero
      ,
      \\
    \label{eq:99}
    \thcstressrhoDiff &= \cstressrho - \pd{\fenergyA}{\henckystrain}.
  \end{align}
\end{subequations}
Equipped with proper boundary and initial conditions the system~\eqref{eq:evolution-equations-system} must be solved for the unknown density~$\rho$, velocity~$\vec{v}$, Hencky strain~$\henckystrain$ and Cauchy stress~$\cstress$. In fact, since $\rho \left(\det \lcg \right)^{\frac{1}{2}}= \rhor$, where $\rhor$ is the material density in the reference configuration, we do not need to solve for $\rho$, because the solution to~\eqref{eq:169} is easy to write down explicitly as $\rho = \frac{\rhor}{\exponential{ \Tr \henckystrain}}$. The material specification enters the governing equations via the density $\rho$, thermodynamic potentials $\fenergyA$ and $\gibbsB$ and the formula for the isotropic fourth order ``rate stiffness'' tensor $\ratestiffnesstensor$. The model is summarised in Summary~\ref{summary:nonlinear-model}, while its linearisation discussed in detail later in Section~\ref{sec:line-disp-relat} is shown in Summary~\ref{summary:linear-model}.

We note that if we set $\ratestiffnesstensor$ equal to zero, then \eqref{eq:70} and \eqref{eq:99} simplify to
\begin{equation}
  \label{eq:92}
  \cstressrho = \pd{\fenergyA}{\henckystrain},
\end{equation}
which is the standard constitutive relation for a hyperelastic/Green elastic solid with the Helmholtz free energy $\fenergyA$.

Finally, since the derivation/formulation of the proposed constitutive relation is based on the standard balance of linear momentum~\eqref{eq:164} and the standard evolution equation for the internal energy~\eqref{eq:24}, we can, following the standard procedures, see, for example, \cite{gurtin.me.fried.e.ea:mechanics}, conclude that a thermodynamically isolated body (no-traction boundary condition, zero heat flux boundary condition) made of a metamaterial described the proposed constitutive relation will preserve its net total energy, that is we shall have
\begin{equation}
  \label{eq:168}
    \dd{}{t}
  \int_{\Omega}
  \left(
    \frac{1}{2}
    \rho
    \vectordot{\vec{v}}{\vec{v}}
    +
    \rho
    \ienergy
  \right)
  \,
  \cvolumee
  =
  0,
\end{equation}
where the internal energy $\ienergy$ is obtained by the Legendre transformation of the generalised Helmholtz free energy $\fenergy$, see~(\ref{eq:68}, supplemented by the purely thermal part thereof.

\begin{summary}[Fully nonlinear model for metamaterial response]
  \label{summary:nonlinear-model}
  \textbf{Material parameters}: Density $\rhor$ in the reference configuration, thermodynamic potentials~$\fenergyA (\temp, \henckystrain)$ (Helmholtz free energy, ``outer/first material'') and $\gibbsB(\temp, \thcstressrhoDiff)$ (Gibbs free energy, ``inner/second material''), isotropic positive definite fourth order ``rate stiffness'' tensor $\ratestiffnesstensor$.
  \smallskip\\
  
  \textbf{Governing equations} in the Eulerian description: For the velocity field~$\vec{v}$, the Hencky strain tensor~$\henckystrain$ and the Cauchy stress tensor~$\cstress$ solve:

  \begin{align*}
    \rho \dd{\vec{v}}{t}
    &=
      \divergence \cstress + \rho \vec{b},
    \\
    \logfid{\overline{\henckystrain}}
    &=
      \gradsym
    \\
    -
    \ratestiffnesstensor
    \logfidd{
      \overline{
        \left(
          \henckystrain
          +
    \pd{\gibbsB}{\thcstressrhoDiff}
        \right)
      }
    }
    +
    \thcstressrhoDiff
    &=
    \tensorzero
    \\
    \thcstressrhoDiff &= \cstressrho - \pd{\fenergyA}{\henckystrain}
  \end{align*}

  \textbf{Notation}: $\gradsym =_{\bydefinition} \frac{1}{2} \left( \nabla \vec{v} + \transpose{\left( \nabla \vec{v} \right)} \right)$, $\cstressrho =_{\bydefinition} \frac{\cstress}{\rho}$, $\rho =_{\bydefinition} \frac{\rhor}{\exponential{\Tr \henckystrain}}$, $\logfid{\overline{\generictensor}}$ -- logarithmic corotational derivative
%
\end{summary}

\begin{summary}[Linearisation of fully nonlinear model for metamaterial response]
  \label{summary:linear-model}
  \textbf{Material parameters}: Density $\rhor$ in the reference configuration, stiffness tensor $\stiffnesstensor$ (``outer/first material'') and weighted inverse to the stiffness tensor $\frac{\rhorB}{\rhor} \inverse{\stiffnesstensorDiff} $ (``inner/second material''), isotropic positive definite fourth order ``rate stiffness'' tensor $\ratestiffnesstensor$.
  \smallskip\\
  
  \textbf{Governing equations}: For the displacement field~$\displacement$ and the Cauchy stress tensor~$\cstress$ solve:
  \begin{align*}
    \rhor
    \ppd{\displacement}{t}
    &=
      \divergence \linstress
    \\
    -
    \ratestiffnesstensor
    \rhor
    \ppd{}{t}
    \left(
    \linstrain
    -
    \frac{\rhorB}{\rhor}
    \inverse{\stiffnesstensorDiff} 
    \left(
      \linstress - \stiffnesstensor \linstrain
    \right)
  \right)
    +
    \left(
      \linstress - \stiffnesstensor \linstrain
    \right)
    =
    \tensorzero
  \end{align*}

  \textbf{Notation}: $\linstrain =_{\bydefinition} \frac{1}{2} \left( \nabla \displacement + \transpose{\left( \nabla \displacement \right)} \right)$
\end{summary}

\section{Dispersion relation}
\label{sec:dispersion-relation}

We now investigate the properties of the model with respect to the wave propagation. In particular, we identify the \emph{dispersion relation} implied by the model~\eqref{eq:evolution-equations-system}. First, we briefly recall the concept of a dispersion relation, and we derive the dispersion relation for the standard (linearised) isotropic elastic solid. We then proceed with the dispersion relation for the linearisation of the model~\eqref{eq:evolution-equations-system}. 

\subsection{Conventional linearised elasticity}
\label{sec:conv-elast-disp}
We assume that the (linearised) stress tensor $\linstress$ is given in terms of the linearised strain tensor $\linstrain$ by the formula
\begin{equation}
  \label{eq:117}
  \linstress
  =
  \left(
    a
    \foidentity
    +
    b
    \tensortensor{\identity}{\identity}
  \right)
  \linstrain,
\end{equation}
that is
$
\linstress
=
b
\left( \Tr \linstrain \right) \identity
+
a \linstrain
$. (The symbol $\identity$ denotes the second order identity tensor, while $\foidentity$ denotes the fourth order identity tensor. In the standard isotropic linearised elasticity theory notation we would have $b =_{\bydefinition} \lambda$ and $a =_{\bydefinition} 2 \mu$, where~$\lambda$ and~$\mu$ are Lam\'e parameters.) If the stress is given by this formula, then the material is described by the evolution equations for the linearised isotropic elastic solid
\begin{subequations}
  \label{eq:118}
  \begin{align}
    \label{eq:119}
    \rho \ppd{\displacement}{t} &= \divergence \linstress, \\
    \label{eq:120}
    \linstress
    &=
    b
    \left( \Tr \linstrain \right) \identity
    +
    a \linstrain.
  \end{align}
\end{subequations}
Here $\displacement$ is the displacement field and $\linstrain =_{\bydefinition} \frac{1}{2} \left( \nabla \displacement + \transpose{\left(\nabla \displacement \right)} \right)$ denotes the linearised strain tensor. The dispersion relation is found using standard manipulations sketched below. We assume that the displacement~$\displacement$ has a plane wave form
\begin{equation}
  \label{eq:166}
  \displacement = \widehat{\displacement} \exponential{ \iunit \left( \vec{k}\cdot \vec{x} - \omega t \right)},
\end{equation}
which essentially means that we take the Fourier transform of the governing equations. Concerning the Fourier image of~\eqref{eq:120} we can write
\begin{equation}
  \label{eq:98}
  \widehat{\linstress}
  =
  \left(
    a
    \foidentity
    +
    b
    \tensortensor{\identity}{\identity}
  \right)
  \widehat{\linstrain}
  .
\end{equation}
Next we observe that
$
  \divergence \linstress
  =
  \left(
    \frac{a}{2} + b
  \right)
  \nabla \divergence \displacement
  +
  \frac{a}{2}
  \Delta \displacement
$,
hence the Fourier transformed version of~\eqref{eq:118} reads
\begin{equation}
  \label{eq:123}
  -
  \rho \omega^2
  \widehat{\displacement}
  =
   -
  \left[
    \left(
      \frac{a}{2} + b
    \right)
    \tensortensor{\vec{k}}{\vec{k}}
    +
    \frac{a}{2}
    \left( \vec{k} \cdot \vec{k} \right)
    \identity
  \right]
  \widehat{\displacement}
  .
\end{equation}
This is the well-known eigenvalue problem for the acoustic tensor. The problem~\eqref{eq:123} has a nontrivial solution~$\widehat{\displacement}$ provided that one of the following conditions is satisfied
\begin{subequations}
  \label{eq:124}
  \begin{align}
    \label{eq:125}
    \rho \omega^2 &= \frac{a}{2} \vec{k} \cdot \vec{k}, \\
    \label{eq:126}
    \rho \omega^2 &= \left(a + b\right) \vec{k} \cdot \vec{k}.
  \end{align}
\end{subequations}
These are the sought dispersion relations. The first equation/condition corresponds to the dispersion relation for the \emph{transversal waves}/shear waves, that is for plane waves~\eqref{eq:166} wherein $\widehat{\displacement} \perp \vec{k}$, the second equation/condition corresponds to the dispersion relation for the \emph{longitudinal waves}/pressure waves, that is for plane waves~\eqref{eq:166} wherein $\widehat{\displacement} \parallel \vec{k}$. The speed of wave propagation $c =_{\bydefinition} \frac{\omega}{\norm{\vec{k}}}$ is then
\begin{subequations}
  \label{eq:127}
  \begin{align}
    \label{eq:128}
    c_{\perp} &=  \sqrt{\frac{a}{2 \rho}}, \\
    \label{eq:129}
    c_{\parallel} &= \sqrt{\frac{a + b}{\rho}},
  \end{align}
\end{subequations}
for the transversal waves and the longitudinal waves respectively. If $b =_{\bydefinition} \lambda$ and $a =_{\bydefinition} 2 \mu$, then we get the standard linearised elasticity formulae for the speed of elastic waves. Note, however, that $a$ and $b$ in the above derivation can be, in principle, arbitrary numbers.

\subsection{Linearised effective constitutive relation}
\label{sec:line-disp-relat}
We now study waves in an elastic solid (metamaterial) described by the model~\eqref{eq:evolution-equations-system}. As common in wave propagation theory, we are interested in the \emph{linearised version} of the model. The standard small displacement gradient theory approximations are
\begin{subequations}
  \label{eq:94}
  \begin{align}
    \label{eq:95}
    \henckystrain & \approx \linstrain, \\
    \label{eq:97}
    \logfid{\overline{\left( \cdot \right)}} & \approx \pd{}{t}\left( \cdot \right),
  \end{align}
\end{subequations}
and similarly for the Hencky strain~$\henckystrainDiff$, $\henckystrainDiff \approx \linstrainDiff$. Furthermore, in the linearised setting the Helmholtz free energy $\fenergyA$ and the Gibbs free energy $\gibbsB$ are approximated by quadratic forms of the appropriate linearised variables. Abusing notation slightly and denoting the quadratic forms approximating the exact Helmholtz free energy and the Gibbs free energy again by $\fenergyA$ and $\gibbsB$, we see that the formulae for the potentials read
\begin{subequations}
\label{eq:c1}
\begin{align}
	\psi_1(\linstrain_1) &= \frac{1}{2\rho_R} \tensordot{\linstrain_1}{\stiffnesstensor \linstrain_1}, \label{eq:c2} \\
	g_2\left (\frac{\linstressDiff}{\rho_{2,R}} \right ) &= -\frac{1}{2\rho_{2,R}} \tensordot{\linstressDiff}{\stiffnesstensorDiff^{-1}\linstressDiff}, \label{eq:c3}
\end{align}
\end{subequations}  
which implies that the linearised versions of constitutive relations for the outer/first continuum, see~\eqref{eq:56}, and the inner/second continuum, see~\eqref{eq:87}, read
\begin{subequations}
  \label{eq:130}
  \begin{align}
    \label{eq:131}
    \linstressA &= \stiffnesstensor \linstrainA,\\
    \label{eq:132}
    \linstressDiff &= \stiffnesstensorDiff \linstrainDiff.
  \end{align}
\end{subequations}
Here, the constant fourth order symmetric positive definite tensors $\stiffnesstensorDiff$ and $\stiffnesstensor$ are the standard isotropic stiffness tensors known from linearised elasticity theory, with possibly different elastic constants defining them. In other words, $\stiffnesstensorDiff$ and $\stiffnesstensor$ represent linear isotropic tensorial functions of the form
$
  c
  \foidentity
  +
  d
  \tensortensor{\identity}{\identity}
$,
where $c$ and $d$ are some constants which are in general different for $\stiffnesstensorDiff$ and~$\stiffnesstensor$. 

We recall that in virtue of the identification~\eqref{eq:55} we have
\begin{equation}
  \label{eq:96}
  \linstrain \equiv \linstrainA,
\end{equation}
thus we can write~\eqref{eq:131} as $\linstressA = \stiffnesstensor \linstrain$. The stress decomposition assumption~\eqref{eq:90} linearises as
\begin{equation}
  \label{eq:100}
  \frac{\linstressDiff}{\rhorB} = \frac{\linstress}{\rhor} - \frac{\linstressA}{\rhor},
\end{equation}
where $\rhor$ denotes the (constant) overall material density in the reference configuration, and $\rhorB$ denotes the (constant) inner/second material density in its reference configuration; recall that $\rho \equiv \rhoA$. Finally, the kinematic assumption~\eqref{eq:88} linearises as
\begin{equation}
 \label{eq:101}
 \linstrainB = \linstrain - \linstrainDiff = \linstrain - \inverse{\stiffnesstensorDiff} \linstressDiff,
\end{equation}
where we have used~\eqref{eq:132}. 

Having collected all partial results, we see that the linearised version of~\eqref{eq:156} reads
\begin{subequations}
  \label{eq:102}
  \begin{align}
    \label{eq:103}
    -
    \ratestiffnesstensor
    \ppd{}{t}
    \left(
    \linstrain
    -
    \inverse{\stiffnesstensorDiff} \linstressDiff
    \right)
    +
    \frac{\linstressDiff}{\rhorB}
    &=
      \tensorzero
      ,
      \\
    \label{eq:104}
    \frac{\linstressDiff}{\rhorB} &= \frac{\linstress - \stiffnesstensor \linstrain}{\rhor}.
  \end{align}
\end{subequations}
Consequently, the full system of linearised governing equations reads
\begin{subequations}
  \label{eq:evolution-equations-system-linearised}
  \begin{align}
    \label{eq:136}
    \rhor
    \ppd{\displacement}{t}
    &=
      \divergence \linstress
      ,
    \\
    \label{eq:138}
    -
    \ratestiffnesstensor
    \rhor
    \ppd{}{t}
    \left(
    \linstrain
    -
    \frac{\rhorB}{\rhor}
    \inverse{\stiffnesstensorDiff} 
    \left(
      \linstress - \stiffnesstensor \linstrain
    \right)
  \right)
    +
    \left(
      \linstress - \stiffnesstensor \linstrain
    \right)
    &=
    \tensorzero,
  \end{align}
\end{subequations}
see also Summary~\ref{summary:linear-model}. This system of equations is the counterpart of the classical linearised elasticity system~\eqref{eq:118}. We again note that if $\ratestiffnesstensor$ is equal to zero, then we recover the classical linearised elasticity case.

We take the Fourier transform of~\eqref{eq:evolution-equations-system-linearised}, and we get
  \begin{subequations}
    \label{eq:linearised-fourier-transformed}
    \begin{align}
      \label{eq:171}
      - \rhor \omega^2 \widehat{\displacement}
      &=
        \widehat{\divergence{\linstress}}, \\
      \label{eq:173}
        \left(
        \foidentity
        -
        \rhorB
        \omega^2
        \ratestiffnesstensor
        \inverse{\stiffnesstensorDiff}
        \right)
      \widehat{\linstress}
      -
      \left(
      \stiffnesstensor
      -
      \rhor
      \omega^2
      \ratestiffnesstensor
      \left(
      \foidentity
      +
      \frac{\rhorB}{\rhor}
      \inverse{\stiffnesstensorDiff}
      \stiffnesstensor
      \right)
      \right)
      \widehat{
      \linstrain
      }
      &
        =
        0.
    \end{align}
  \end{subequations}
Note that in virtue of identity 
\begin{equation}
  \label{eq:116}
  \inverse{
    \left(
      u
      \foidentity
      +
      v
      \tensortensor{\identity}{\identity}
    \right)
  }
  =
  \frac{1}{u}
  \left(
    \foidentity
    -
    \frac{v}{u + 3v}
    \tensortensor{\identity}{\identity}
  \right)
\end{equation}
we are in principle able to explicitly calculate all the inverses of fourth order tensors in~\eqref{eq:173} and rewrite~\eqref{eq:173} in the form identical to the Fourier transformed version of~\eqref{eq:117}, see also~\eqref{eq:98}, which is the form known from the classical linearised elasticity theory. \emph{However, the coefficients $a$ and $b$ would now depend on the wave frequency $\omega$}. Thus, in the Fourier space, the proposed model is mathematically equivalent to a classical isotropic elastic solid with frequency dependent material moduli.

We elaborate this observation for a particularly simple version of~\eqref{eq:173}. For simplicity, we assume 
\begin{subequations}
  \label{eq:140}
  \begin{align}
    \label{eq:141}
    \ratestiffnesstensor &=_{\bydefinition}  c_{\text{rate}} \stiffnesstensor, \\
    \label{eq:142}
    \stiffnesstensorDiff &=_{\bydefinition} c_{\text{diff}} \stiffnesstensor,
  \end{align}
\end{subequations}
that is if the fourth order tensors $\ratestiffnesstensor$ and $\stiffnesstensorDiff$ are just scalar multiples of $\stiffnesstensor$, wherein $\stiffnesstensor$ has the form
\begin{equation}
  \label{eq:69}
  \stiffnesstensor
  =_{\bydefinition}
    \left(
    2 \mu
    \foidentity
    +
    \lambda
    \tensortensor{\identity}{\identity}
  \right).
\end{equation}
This means that the outer/first continuum is just a linearised isotropic elastic solid with Lam\'e parameters $\mu$ and $\lambda$. Then the Fourier transformed effective constitutive relation~\eqref{eq:173} reduces to
\begin{equation}
  \label{eq:143}
  \left(
    1
    -
    \rhorB
    \frac{c_{\text{rate}}}{c_{\text{diff}}}
    \omega^2
  \right)
  \widehat{\linstress}
  -
  \left(
    1
    -
    \rhor
    c_{\text{rate}}
    \left(
      1
      +
      \frac{\rhorB}{\rhor}
      \frac{1}{c_{\text{diff}}}
    \right)
    \omega^2
  \right)
  \stiffnesstensor
  \widehat{
    \linstrain
  }
  =
  \tensorzero
  ,
\end{equation}
which can be rewritten as
\begin{equation}
  \label{eq:121}
  \widehat{\linstress}
  =
  \frac{
    1
    -
    \rhor
    c_{\text{rate}}
    \left(
      1
      +
      \frac{\rhorB}{\rhor}
      \frac{1}{c_{\text{diff}}}
    \right)
    \omega^2
  }
  {
    1
    -
    \rhorB
    \frac{c_{\text{rate}}}{c_{\text{diff}}}
    \omega^2
  }
  \stiffnesstensor
  \widehat{\linstrain}
  .
\end{equation}
Consequently, we see that we are indeed formally in the same position as in the classical linearised elasticity setting, see~\eqref{eq:98} in Section~\ref{sec:conv-elast-disp}, but with a substantial twist. Unlike in~Section~\ref{sec:conv-elast-disp}, the coefficients $a$ and $b$ in~\eqref{eq:98} are now frequency dependent. Nevertheless, the linear stress--strain relation with frequency dependent coefficients~\eqref{eq:121} can be algebraically manipulated exactly in the same manner as in Section~\ref{sec:conv-elast-disp}, and upon substituting~\eqref{eq:121} into~\eqref{eq:171} we end up with the equation
\begin{equation}
  \label{eq:144}
  -
  \rhor \omega^2
  \widehat{\displacement}
  =
  -
    \frac{
    1
    -
    \rhor
    c_{\text{rate}}
    \left(
      1
      +
      \frac{\rhorB}{\rhor}
      \frac{1}{c_{\text{diff}}}
    \right)
    \omega^2
  }
  {
    1
    -
    \rhorB
    \frac{c_{\text{rate}}}{c_{\text{diff}}}
    \omega^2
  }
  \left[
    \left(
      \mu + \lambda
    \right)
    \tensortensor{\vec{k}}{\vec{k}}
    +
    \mu
    \left( \vec{k} \cdot \vec{k} \right)
    \identity
  \right]
  \widehat{\displacement}
  ,
\end{equation}
which can be further rewritten as
\begin{equation}
  \label{eq:145}
  -\rhor
  \underbrace{
  \left(
    1
    +
    \frac{
      \rhor c_{\text{rate}} \omega^2
    }
    {
      1
      -
      \rhor
      c_{\text{rate}}
      \left(
        1
        +
        \frac{\rhorB}{\rhor}
        \frac{
          1
        }
        {
          c_{\text{diff}}
        }
      \right)
      \omega^2
    }
  \right)
  }_{e(\omega)}
  \omega^2
  \widehat{\displacement}
  =-
  \left[
    \left(
      \mu + \lambda
    \right)
    \tensortensor{\vec{k}}{\vec{k}}
    +
    \mu
    \left( \vec{k} \cdot \vec{k} \right)
    \identity
  \right]
  \widehat{\displacement}.
\end{equation}
The speed of wave propagation $c =_{\bydefinition} \frac{\omega}{\norm{\vec{k}}}$ is easy to read from~\eqref{eq:145} in terms of the correction factor~$e(\omega)$. (See the similar discussion in the classical linearised elasticity case, especially the discussion following equation~\eqref{eq:123}.) The formulae for the (phase) speed of transversal/shear and longitudinal/pressure waves read
\begin{subequations}
  \label{eq:170}
  \begin{align}
    c_{\perp}^2 &=  \frac{\mu}{\rhor} \frac{1}{e(\omega)}, \\
    \label{eq:129}
    c_{\parallel}^2 &= \frac{\lambda + 2 \mu}{\rhor} \frac{1}{e(\omega)},
  \end{align}
\end{subequations}
which can be compared with the classical linearised elasticity formulae, see~\eqref{eq:127}, which yield $c_{\perp}^2 =  \frac{\mu}{\rhor}$ and~$c_{\parallel}^2 = \frac{\lambda + 2 \mu}{\rhor}$.

Finally, we note that equation~\eqref{eq:145} has the same structure as the equation used in metamaterial models based on the concept of frequency dependent density, see~\cite[Eq. 20]{rizzi.g.d’agostino.mv.ea:from}, which are the models that lead to the ``negative mass density'' inconsistency. The proposed model~\eqref{eq:evolution-equations-system-linearised} can thus reproduce all the features as popular metamaterial models, but \emph{without} the need to introduce the physically dubious concept of motion dependent density.

\section{Conclusion}
\label{sec:conclusion}

We have proposed a thermodynamically based approach for constructing effective rate-type constitutive relations~\eqref{eq:137} describing fully nonlinear (finite deformations) behaviour of metamaterials. The effective constitutive relation is formulated as a second-order in time rate-type Eulerian constitutive relation between the Cauchy stress tensor, the Hencky strain tensor and their logarithmic corotational derivatives. There is no need to introduce additional quantities or concepts such as ``micro-level deformation'' and so forth---the macroscopic stress and strain fields are the only quantities necessary for the formulation of the constitutive relation and for the description of metamaterial behaviour.

The effective constitutive relation has been derived in analogy to the paradigmatic discrete mass-in-mass system introduced by~\cite{huang.hh.sun.ct.ea:on}, and the potentials $\fenergyA$ and $\gibbsB$ that appear in the constitutive relation are loosely identified with the Helmholtz free energy and the Gibbs free energy of the metamaterial components. In principle, these potentials might be chosen from a wide catalogue of classical potentials known for elastic solids, see, for example, \cite{mihai.la.goriely.a:how} for the classical Helmholtz free energy approach, and, for example, \cite{muliana.a.rajagopal.kr.ea:determining}, \cite{prusa.v.rajagopal.kr.ea:gibbs} and \cite{bertoti.e:non-linear} for the dual/Gibbs free energy approach.

Since the effective constitutive relation is from the very beginning formulated in the \emph{fully nonlinear setting} for finite deformations, it might provide a suitable framework for any future study of inherently nonlinear phenomena in metamaterials. One such example might be the classical Poynting effect, see, for example, \cite{horgan.co.murphy.jg:poynting}, and its possible analogue in the dynamical response of metamaterials. Similarly, since the effective constitutive relation has been formulated in the full thermodynamic setting, it might provide a suitable framework for any future study of complex nonlinear thermomechanical phenomena in metamaterials. 

The linearisation of the proposed fully nonlinear system effectively leads, in Fourier space, to the same constitutive relation as those commonly used in theories based on the concept of frequency dependent density and/or stiffness. This opens up the possibility of parameter identification using the same procedures as those used in the case of frequency dependent density and/or stiffness models. The proposed approach however works with constant---that is motion independent---material properties, which is clearly more convenient from the physical point of view. Furthermore, the proposed linearised system of governing equations is particularly simple both in Fourier space as well as in physical space. We also note that the linearisation that leads to the popular frequency dependent density and/or stiffness models is based on a very special choice of material parameters in the newly proposed model, namely all the (generalised) stiffness tensors differ at most by a scalar factor. This is not the only possibility, and more general choices of (generalised) stiffness tensors might lead to more involved wave propagation phenomena. However, a detailed analysis of this parameter choice is beyond the scope of the current work and warrants future investigation.

The same need for future investigation holds regarding numerical simulations based on the proposed model. (Even in the linear setting.) Here the already existing micromorphic type metamaterial models have a clear advantage as numerical simulations based on micromorphic type models are already existing, see, for example, \cite{perez-ramirez.la.rizzi.g.ea:multi-element} and references therein. Only a particular implementation can show whether the just proposed approach provides a viable alternative to micromorphic models regarding the numerical simulation of metamaterial behaviour.    

Finally, if we accept that the rate-type constitutive relation~\eqref{eq:137} describes an elastic solid---in the sense that it describes a solid material that does not dissipate the energy in mechanical processes and that allows mutual conversion of the kinetic and the ``stored'' energies---then we must reconsider our approach to the general description of elastic materials. Clearly, the proposed rate-type constitutive relations do not fall into the class of ``potential'' based elasticity theories (hyperelasticity) embodied by the classical Doyle--Ericksen formula, see, for example, \cite{truesdell.c.noll.w:non-linear*1}, or (algebraic) implicit type theories, see~\cite{rajagopal.kr:on*3}, or classical \emph{first-order} in time rate-type (hypoelastic) theories, see \cite{truesdell.c:hypo-elasticity}, \cite{noll.w:on}, \cite{bernstein.b.ericksen.jl:work}, \cite{bernstein.b:relations,bernstein.b:hypo-elasticity}, \cite{xiao.h.bruhns.ot.ea:hypo-elasticity}, \cite{leonov.ai:on*1} and \cite{bernstein.b.rajagopal.k:thermodynamics} to name a few. Thus, the \emph{second-order} rate-type constitutive relations of the type~\eqref{eq:137} simply embody a new class of constitutive relations characterising elastic solids.